\documentclass[a4paper, 12pt]{article}
\usepackage{arxiv}

\usepackage{microtype}
\usepackage{graphicx}
\usepackage{wrapfig}
\usepackage{enumitem}
\usepackage{cite}
\usepackage{amsmath,bm}
\usepackage{authblk}
\usepackage{algorithm}
\usepackage{algorithmic}
\usepackage{chngcntr}

\title{\Large{\textbf{Learned SPARCOM: Unfolded Deep Super-Resolution Microscopy}}} 

\author[1]{Gili Dardikman-Yoffe$^*$}
\author[1]{Yonina C. Eldar}

\affil[1]{Faculty of Mathematics and Computer Science, Weizmann institute of science, Israel
*Corresponding author: gili.dardikman@weizmann.ac.il}

\date{}                    
\begin{document}

\maketitle

\begin{abstract}
The use of photo-activated fluorescent molecules to create long sequences of low emitter-density diffraction-limited images enables high-precision emitter localization, but at the cost of low temporal resolution. 
We suggest combining SPARCOM, a recent high-performing classical method, with model-based deep learning, using the algorithm unfolding approach, to design a compact neural network incorporating domain knowledge. 
Our results show that we can obtain super-resolution imaging from a small number of high emitter density frames without knowledge of the optical system and across different test sets using the proposed learned SPARCOM (LSPARCOM) network.
We believe LSPARCOM can pave the way to interpretable, efficient live-cell imaging in many settings, and find broad use in single molecule localization microscopy of biological structures.
\end{abstract}

\section{Introduction}
Breakthroughs in the field of chemistry have enabled surpassing the classical optical diffraction limit, yielding significantly higher resolution, by utilizing photo-activated fluorescent molecules. In the approach called 2-D single-molecule localization microscopy (SMLM) \cite{PALM, STORM}, a sequence of diffraction-limited images, produced by a sparse set of emitting fluorophores with minimally overlapping point-spread functions (PSFs) is acquired, allowing the emitters to be localized with high precision by relatively simple post-processing. This enables imaging subcellular features and organelles within biological cells with unprecedented resolution. The low emitter density concept, which is manifested in some of the most promising imaging methods, including PALM \cite{PALM} and STORM \cite{STORM}, requires lengthy imaging times to achieve full coverage of the imaged specimen on the one hand, and minimal overlap between PSFs on the other. Thus, this concept in its classical form has low temporal resolution, limiting its application to slow-changing specimens and precluding more general live-cell imaging. 

To circumvent the long acquisition periods required for SMLM methods, a variety of techniques have emerged, which enable the use of a smaller number of frames for reconstructing the 2-D super-resolved image 
\cite{FALCON, CS-STORM, SOFI, SPARCOM, SPARCOM2, ANNA-PALM, DEEP-STORM}. These techniques take advantage of prior information regarding either the optical setup, the geometry of the sample, or the statistics of the emitters. 
In \cite{FALCON, CS-STORM} the authors consider frame-by-frame recovery by relying on sparse coding approaches \cite{CS_BOOK, SAMPL_BOOK}.
To utilize the statistical information in the full temporal stack of images, super-resolution optical fluctuation imaging (SOFI) uses high-order statistical analysis (cumulants) of the temporal fluctuations \cite{SOFI}, enabling reducing the effective PSF size according to the squared root of the order of the cumulant. In practice, however, the use of statistical orders higher than two is limited due to signal-to-noise ratio (SNR), dynamic range expansion, and temporal resolution considerations, leaving the spatial resolution practically offered by SOFI significantly lower than PALM and STORM. 
Solomon et al. have recently suggested combining the ideas of sparse recovery and SOFI, leading to a sparsity-based approach for super-resolution microscopy from correlation information of high emitter-density frames, dubbed SPARCOM \cite{SPARCOM, SPARCOM2}. SPARCOM utilizes sparsity in the correlation domain, while assuming that the blinking emitters are uncorrelated over time and space. This was shown to increase the number of detected locations of the sources compared with sparse recovery performed on the signal itself \cite{SparseSupport}, and thus improve both temporal and spatial resolution compared to counterpart methods. 
Nevertheless, SPARCOM requires heuristic adjustment of optimization parameters and prior knowledge of the optical system PSF.

The past decade has seen an explosion in the use of deep learning algorithms \cite{DL0, DL1, DL2, DL3} across all areas of science. It is thus natural to consider whether temporal resolution can be improved using deep learning techniques. 
One such approach is ANNA-PALM \cite{ANNA-PALM}, which reconstructs high-quality images from undersampled localization microscopy data, obtained by a small number of low-emitter-density frames, utilizing the implicit structural redundancy of most biological images. In this setting, the network receives as input a small number of frames that do not contain PSF overlaps, such that the emitters can be localized with high fidelity, but without supplying full coverage of the specimen; the completion of the missing emitters allowing deduction of the full structure of the specimen is achieved using prior training on pairs of sparse and dense PALM images with structures similar to those in the images to be reconstructed. As can be expected, ANNA-PALM fails in the absence of statistical redundancies between molecular localizations, and does not generalize well to structures different than those it was trained on. 
An alternative deep learning technique overcoming some of these shortcomings is Deep-STORM \cite{DEEP-STORM}; unlike ANNA-PALM it takes high-emitter-density frames as inputs, and uses an encoder-decoder neural network architecture to restore a super-resolved image, relying on prior information regarding the optical setup which it gets from the training set. In contrast to ANNA-PALM, Deep-STORM generalizes very well to PALM images with different structures, not requiring structure-targeted training; however, as demonstrated in the results section, it does not generalize well to test data with different imaging parameters (including camera base level, photoelectrons per A/D count, PSF, emitter FWHM range, emitter intensity range and mean photon background) than those of the training data. Furthermore, it has many parameters to train, and does not have an interpretable structure, which can be valuable particularly in biological settings.

Thus, currently, super-resolution reconstruction techniques either rely on classical algorithms (such as SPARCOM \cite {SPARCOM}), which have low dependency on the type of  input data as long as it fits the prior, but require explicit knowledge of the optical setup and are highly dependent on tuning the optimization parameters; or standard deep-learning based methods (such as ANNA-PALM \cite {ANNA-PALM} and Deep-STORM \cite {DEEP-STORM}), which have very strong dependencies on the training data: ANNA-PALM requires the test set to be identical to the training set in terms of structure, and Deep-STORM - in terms of imaging parameters. 
Deep-STORM and ANNA-PALM both use generic network architectures that are not easily interpretable, and require many layers and parameters to yield good performance. Here, our goal is to develop an interpretable, efficient, deep network that will train from even a single field of view (FoV), generalize well, not rely on explicit knowledge of the optical setup, and not require fine-tuning of optimization parameters.

A decade ago, Gregor and Lecun suggested deep algorithm unfolding (unrolling) as a strategy for designing neural networks based on domain knowledge \cite{LISTA}. 
In this approach, the network architecture is tailored to a specific problem, based on a well-founded iterative mathematical formulation for solving the problem at hand \cite{UNFOLD_REV}. This leads to increased convergence speed and accuracy with respect to the standard, iterative solution, and interpretability and robustness relative to a black-box large-scale neural network \cite{UNFOLD_REV, DEEP_UNFOLD}.
In their seminal work, Gregor and LeCun unfolded the Iterative Shrinkage-Thresholding Algorithm (ISTA) for sparse coding \cite{ISTA1, ISTA2} into a learned ISTA (LISTA) network, and demonstrated that propagating the data through only 10 blocks is equivalent to running the iterative algorithm for 200 iterations, without requiring fine-tuning of any parameters. In recent years, the concept of algorithm unfolding has been applied to many different problems, including, among others, single-image-super-resolution (deblurring) \cite{DEBLUR0, DEBLUR}, image denoising and image inpainting  \cite{LISTA_CONV}, ultrasound localization microscopy \cite{US_SR}, ultrasound clutter suppression \cite{CORONA}, and multi-channel source separation \cite{SOURCE_SEP}.

In this paper, we utilize for the first time the deep algorithm unfolding concept to transform the correlation-domain sparsity-based approach suggested by Solomon et al. \cite{SPARCOM, SPARCOM2} into a simple deep learning, parameter-free framework, dubbed LSPARCOM (learned SPARCOM), which we train on a single FoV. 
Our method is robust, generalizes well, is interpretable, and requires only a small number of layers, without relying on explicit knowledge of the optical setup or requiring fine-tuning of optimization parameters.
We compare the resultant LSPARCOM network both to the iterative algorithm it is based on, SPARCOM \cite{SPARCOM}, and to the leading classical deep-learning based localization method for high emitter-density frames, Deep-STORM \cite{DEEP-STORM}, by testing them on simulated and experimental data. The results show that LSPARCOM is able to yield comparable or superior results to those obtained by SPARCOM with no heuristic parameter determination or explicit knowledge of the PSF; in addition, they demonstrate that LSPARCOM generalizes better than Deep-STORM, and is able to yield excellent reconstruction with as few as 25 frames. This is made possible by the use of a small, parameter-efficient neural network which uses a model-based framework, thus not requiring data-specific training - neither in terms of structure nor in terms of imaging parameters. We also compare runtimes, showing that the small network size ($143 \times$ less parameters than Deep-STORM) contributes to significantly faster execution times relative to both SPARCOM and Deep-STORM.
This paves the way to true live-cell imaging using a compact, efficient, interpretable deep network that can generalize well to unseen settings, train from a single FoV, and require a very small number of frames over time.
 
The rest of the paper is organized as follows. 
In Section 2 we formulate the super-resolution optical problem. 
In Section 3, we explain SPARCOM, the method we base our network on. 
Unfolding of SPARCOM to LSPARCOM is performed in Section 4, where we also describe the implementation details.
Simulation and experimental results showing the performance of LSPARCOM and comparing it to previously proposed techniques are provided in Section 5. 
Finally, we discuss both the impact and limitations of our approach in Section 6. 

Throughout the paper, $t$ represents time, $x$ represents a scalar, $\textbf{x}$ denotes a vector, $\textbf{X}$ a matrix, and $\textbf{I}_{N\times N}$ is the $N\times N$ identity matrix. Subscript $x_l$ denotes the $l$'th element of  $\textbf{x}$, and  $\textbf{x}_l$  is the $l$'th column of $\textbf{X}$. Superscript $\textbf{x}^{(k)}$ represents $\textbf{x}$ at iteration $k$, and $\textbf{A}^T$ is the transpose of  $\textbf{A}$.

\section{Problem formulation}
In the SMLM setting, we aim to recover a single $N\times N$ high resolution image $\textbf{X}$, corresponding to the locations of the emitters on a fine grid, from a set of $T$ low-resolution $M\times M$ frames $f(t)$, where $M<N$. This enables recovery of fine features, smaller than the size of a single camera pixel. 

The relation between the low-resolution $M\times M$ frame $f(t)$ acquired at time $t$ and the location of the emitters on the high resolution grid $\textbf{X}$ can be formulated considering the blinking and diffraction phenomena as follows \cite{SPARCOM, SPARCOM2}:

\begin{equation} 
f[m\Delta_L,n\Delta_L,t] =\sum_{i=0}^{N-1} \sum_{l=0}^{N-1}u[m\Delta_L-i\Delta_H,n\Delta_L-l\Delta_H]s_{il}(t).
\end{equation}
Here $\Delta_L$ is the spacing of the low-resolution grid and $\Delta_H$ is the spacing of the high-resolution grid such that $M\Delta_L = N\Delta_H$, $u[m\Delta_L, n\Delta_L]$ , $m, n = [0, ... M-1] $  is the low-resolution sampled PSF, and $s_{il}(t)$ represents the temporal fluctuation of the emitter in pixel $[i,l]$ on the high resolution grid. 
We stack the $N\times N$ matrix obtained by all entries ($i,l=0...N-1$) of $s_{il}(t)$, yielding the vector $\textbf{s}(t)$. Assuming only $L$ emitters exist on the $N\times N$ grid, there are no more than $L$ out of $N^2$ locations in vector $\textbf{s}(t)$ which have non-zero values, meaning that it is at least $L$ sparse for all times. 
The actual values in $\textbf{s}(t)$ are not as important as its support, meaning which locations have non-zero values, as these indicate locations with emitters on the high resolution grid. For each time $t$ the support of $\textbf{s}(t)$ is identical to the support of the corresponding high resolution image, such that they are interchangeable.
Accordingly, if we sum $\textbf{s}(t)$ for all times and reshape it as an image, we find that the support of the result is identical to the support of the final desired high resolution image $\textbf{X}$. 

For each $t$, (1) can be formulated using the following matrix multiplication:
\begin{equation} 
\textbf{y}(t)= \textbf{A}\textbf{s}(t).
\end{equation}
Here, \textbf{y}(t) is the $M^2$ vector stacking of the $M \times M$ frame $f(t)$ taken at time $t$, and \textbf{A} is the $M^2 \times N^2$ measurement matrix, in which each column is an $M^2$ long vector corresponding to an $M$  $\times M$ image of the PSF shifted to a different location on the high resolution grid. Thus, for each time $t$ the non-zero elements of $\textbf{s}(t)$ multiply the corresponding atoms in \textbf{A}, producing the linear combination which describes the recorded image \textbf{y}(t). 
Equation (2) defines a single measurement vector (SMV) model, where we seek to find the locations of the emitters on the high resolution grid $\textbf{s}(t)$ for each frame independently, given the recorded frame \textbf{y}(t) and the measurement matrix \textbf{A}. 
The SMV model can be combined with a sparse prior on the high-resolution reconstruction $\textbf{s}(t)$ \cite{CS_BOOK}, relying on the fact that emitters are sparsely distributed - a reasonable assumption considering the finer grid used for reconstruction relative to recording. This is the approach used in \cite{FALCON, CS-STORM}, where the authors consider frame-by-frame recovery relying on sparse coding techniques.

Assuming the entire set of $T$ low-resolution $M\times M$ frames $f(t)$ is acquired without the underlying sample changing, but only with blinking of the emitters, all high resolution reconstructions $\textbf{s}(t)$ share the same support (non-zero locations). This additional information can be used for improved recovery, by considering the multiple measurement vector (MMV) model \cite{CS_BOOK, MMV_PAPER}, where reconstruction is achieved jointly for all measurements. This problem can be formulated as
\begin{equation} 
\textbf{Y}= \textbf{A}\textbf{S}.
\end{equation}
Here, $\textbf{Y}$ is an $M^2\times T$ matrix, where each column is the vector stacking \textbf{y}(t) of a single $M \times M$ frame, \textbf{S} is an $N^2\times T$ matrix, where each column is the $L$-sparse vector stacking $\textbf{s}(t)$ of a single $N\times N$ high-resolution sparse frame, and \textbf{A} is the same $M^2 \times N^2$ measurement matrix. For each column of $\textbf{S}$, the non-zero elements multiply the corresponding atoms in \textbf{A}, producing the linear combination which describes the corresponding column in \textbf{Y}. Given \textbf{Y} and \textbf{A}, we aim to recover the matrix \textbf{S}; the sum of its columns reshaped as a matrix yields the final desired high resolution image $\textbf{X}$.

SOFI \cite{SOFI} uses prior knowledge regarding the statistics of the blinking emitters. In the following section we present SPARCOM \cite{SPARCOM}, which suggests solving the MMV compressed sensing problem using additional prior information regarding the statistics of the emitters, yielding a correlation-based MMV approach.

\section{SPARCOM: sparsity-based super-resolution microscopy from correlation information}

In SPARCOM \cite{SPARCOM, SPARCOM2}, Solomon et al. assume that emissions by different emitters are uncorrelated over time and space, providing further prior information to exploit for solving the compressed sensing MMV problem given in (3). 

To derive SPARCOM, we first subtract from each column in $\textbf{Y}$ its temporal mean, yielding a new matrix which we still denote as $\textbf{Y}$ for simplicity. 
Then, we obtain the covariance matrices of $\textbf{Y}$ and $\textbf{S}$ by multiplying each side of (3) by its transpose, yielding
\begin{equation} 
\textbf{Y}\textbf{Y}^T = [\textbf{A}\textbf{S}] [\textbf{A}\textbf{S}]^T = \textbf{A}\textbf{S}\textbf{S}^T\textbf{A}^T.
\end{equation}
Denoting $\textbf{R}_y \overset{\Delta}{=} \textbf{Y}\textbf{Y}^T $ and $\textbf{R}_s \overset{\Delta}{=} \textbf{S}\textbf{S}^T $ we have
\begin{equation} 
\textbf{R}_y= \textbf{A}\textbf{R}_s\textbf{A}^T.
\end{equation}
The left-hand side of (5) is calculated empirically from the data, whereas \textbf{A} is calculated from the PSF, by fitting the resolution-limited image of the PSF to a Gaussian, plotting it on the high-resolution grid, and shifting its center to all possible locations. The covariance matrix $\textbf{R}_s$ is the unknown that needs to be recovered. However, since the emitters are assumed to be uncorrelated, $\textbf{R}_s$ is a diagonal matrix, such that the number of unknowns is $N^2$ rather than $N^4$. This is not the case for $\textbf{R}_y$, since $\textbf{Y}$ accounts for diffraction, so that the values in different pixels are correlated according to the PSF. 

The diagonal of $\textbf{R}_s$, which we denote by $\textbf{x}$, represents the variance of the emitter fluctuation on a high-resolution grid.
The actual values in the vector $\textbf{x}$ are not as important as its support, which indicate locations with emitters on the high resolution grid. The support of $\textbf{x}$ is identical to the support of the corresponding vector-stacked high resolution image obtained from all frames.
Therefore, recovering $\textbf{x}$ and reshaping it as a matrix yields the desired $\textbf{X}$.
Since we assume that out of $N^2$ pixels there are only $L$ emitters, $\textbf{x}$ is $L$-sparse.
Note that this assumption of sparsity does not mean that the emitters should be isolated from each other in the classical sense of single-molecule localization microscopy, which mandates a lack of overlap between the images formed by adjacent emitters; but only that the total number of pixels consisting of emitters in the high-resolution grid is substantially smaller than the total number of pixels. 

The equivalence of the variance and emitter location is only true as long as all emitters blink. An emitter with a constant value throughout the recording process will have a zero value in the variance vector $\textbf{x}$. However, we assume that no emitter maintains a constant value throughout the recording, such that the equivalence holds.

Reformulating (5) given that $\textbf{R}_s = $ diag$\{\textbf{x}\}$ yields:
\begin{equation} 
\textbf{R}_y = \sum_{l=1}^{N^2}\textbf{a}_l\textbf{a}_l^Tx_l,
\end{equation}
where $x_l$ is the $l$th entry of $\textbf{x}$ and $\textbf{a}_l$ is the $l$th column of $\textbf{A}$.
The high resolution image $\textbf{X}$ can then be found by solving the following convex optimization problem for its vector-stacking $\textbf{x}$:
\begin{equation} 
\!\min_{x\geq0}   \lambda\|\textbf{x}\|_1 + f(\textbf{x})
\end{equation}
with
\begin{equation} 
f(\textbf{x}) = \frac{1}{2}\|\textbf{R}_y - \sum_{l=1}^{N^2}\textbf{a}_l\textbf{a}_l^Tx_l \|_F^2, 
\end{equation}
where  $\lambda \geq 0$ is the regularization parameter (determined heuristically) and $\|\cdot\|_F$ denotes the Frobenuis norm. In this formulation, the $L_1$-regularizer $\|\cdot\|_1$ promotes sparsity of $\textbf{x}$, whereas $f(\textbf{x})$ enforces consistency with the physical model. 
The optimization problem defined above can be solved iteratively using ISTA \cite{ISTA1, ISTA2}, leading to Algorithm 1. Note that in \cite{SPARCOM, SPARCOM2} the faster version (FISTA) is used \cite{FISTA, FISTA2}, yet for the sake of the unfolded algorithm, which will be presented in the following section, we are only interested in the basic framework and not in optimizing the convergence rate of the iterative counterpart. 

\begin{algorithm}
\caption{SPARCOM via ISTA for minimizing (7)}
\begin{algorithmic}
\STATE \textbf{Input:} $\textbf{v}, \textbf{M}, L_f, \lambda \geq 0, K_{MAX} $
\STATE \textbf{Initialize:} $\textbf{x}^{(0)} = 0$
\STATE \textbf{For} $k = 0:K_{MAX} -1$
\item $ \textbf{x}^{(k+1)} = T^+_{\frac{\lambda}{L_f}}[\frac{1}{L_f}\textbf{v} + (\textbf{I}_{N^2\times N^2}-\frac{1}{L_f}\textbf{M})\textbf{x}^{(k)}]$
\STATE \textbf{end} 
\STATE \textbf{Output:} $\textbf{x}^{(K_{MAX})}$ 
\end{algorithmic}
\end{algorithm}

The inputs to Algorithm 1 include the regularization parameter $\lambda \geq 0$, the number of iterations $K_{MAX}$, the vector $\textbf{v}$, which stores the information regarding the recorded data
\begin{equation} 
\textbf{v} = [\textbf{a}_1^T\textbf{R}_y\textbf{a}_1,... , \textbf{a}_{N^2}^T\textbf{R}_y\textbf{a}_{N^2}]^T,
\end{equation}
 as well as two additional setup constants, $\textbf{M}$ and $L_f$, where
 \begin{equation} 
\textbf{M} = |\textbf{A}^T\textbf{A}|^2,
\end{equation} 
and $L_f$ is the Lipschitz constant of the gradient of (8), and is equal to the maximum eigenvalue of $\textbf{M}$.
The variables $\textbf{M}$ and $\textbf{v}$ are obtained by calculating the gradient of (8); differentiating (8) with respect to $\textbf{x}$ results in $\nabla$$f(\textbf{x}) = \textbf{M}\textbf{x}-\textbf{v}$.
The absolute value in (10) is taken element wise.

The operator $T^+_\alpha(\cdot)$ in Algorithm 1 is the positive soft thresholding operator with parameter $\alpha$, which is equal to the shifted rectified linear unit (ReLU), defined by:
\begin{equation} 
T^+_\alpha(x) \overset{\Delta}{=} ReLU(x-\alpha) = \text{max}\{x-\alpha, 0\}.
\end{equation}
In (11) $x$ is scalar; when applied to vectors and matrices $T^+_{\alpha}(\cdot)$ operates element-wise.
The use of the soft thresholding operator is derived from (7), as it is the proximal operator of the $L_1$-norm. The use of the positive soft thresholding operator (rather than a standard soft thresholding operator) is a result of the problem at hand; since $\textbf{x}$ represents the variance of the emitters it is necessarily non-negative. 

While SPARCOM yields excellent results \cite{SPARCOM}, it requires knowledge of the optical PSF for calculation of the measurement matrix $\textbf{A}$, as well as heuristic choice of a regularization parameter $\lambda$, determined by trial and error.
In the following section we lay out the unfolding process of SPARCOM to LSPARCOM, enabling end-to-end learning of optimal parameters from training data with no prior knowledge of the optical setup.

\section{Learned SPARCOM (LSPARCOM)}

SPARCOM is based on a rigorous mathematical approach; as such, it requires both prior knowledge of the PSF of the optical setup for the calculation of the measurement matrix, which is not always available, and a wise choice of regularization factor $\lambda$, which is generally done heuristically. Here we show how we can overcome these shortcomings by learning these from data using an algorithm unfolding approach.
 
The idea at the core of deep algorithm unfolding, as first suggested by Gregor and Lecun \cite{LISTA}, is using the algorithmic framework to gain interpretability and domain knowledge, while inferring optimal parameters from the data itself. In this strategy, the design of a neural network architecture is tailored to the specific problem, based on a well-founded iterative mathematical formulation for solving the problem. 

Figure 1 is a schematic illustration of the unfolding of Algorithm 1 into a deep network with 10 layers, which we term LSPARCOM. Each step of the unfolding process is detailed in the following subsections, explaining the choice of input, the replacement of multiplication layers with convolutional filters, the choice of activation function and the end-to-end parameter learning process. 
Algorithm 2 presents LSPARCOM at inference time in more detail, mentioning the final layer which was excluded from Fig. 1 - a trainable scaling layer at the output $s$, allowing simple adjustment of the scale relative to the ground truth. The trainable parameters are learned from the data during training, as explained in Section 4.4.

\begin{figure}[h] 
\centering
\includegraphics[scale=0.2]{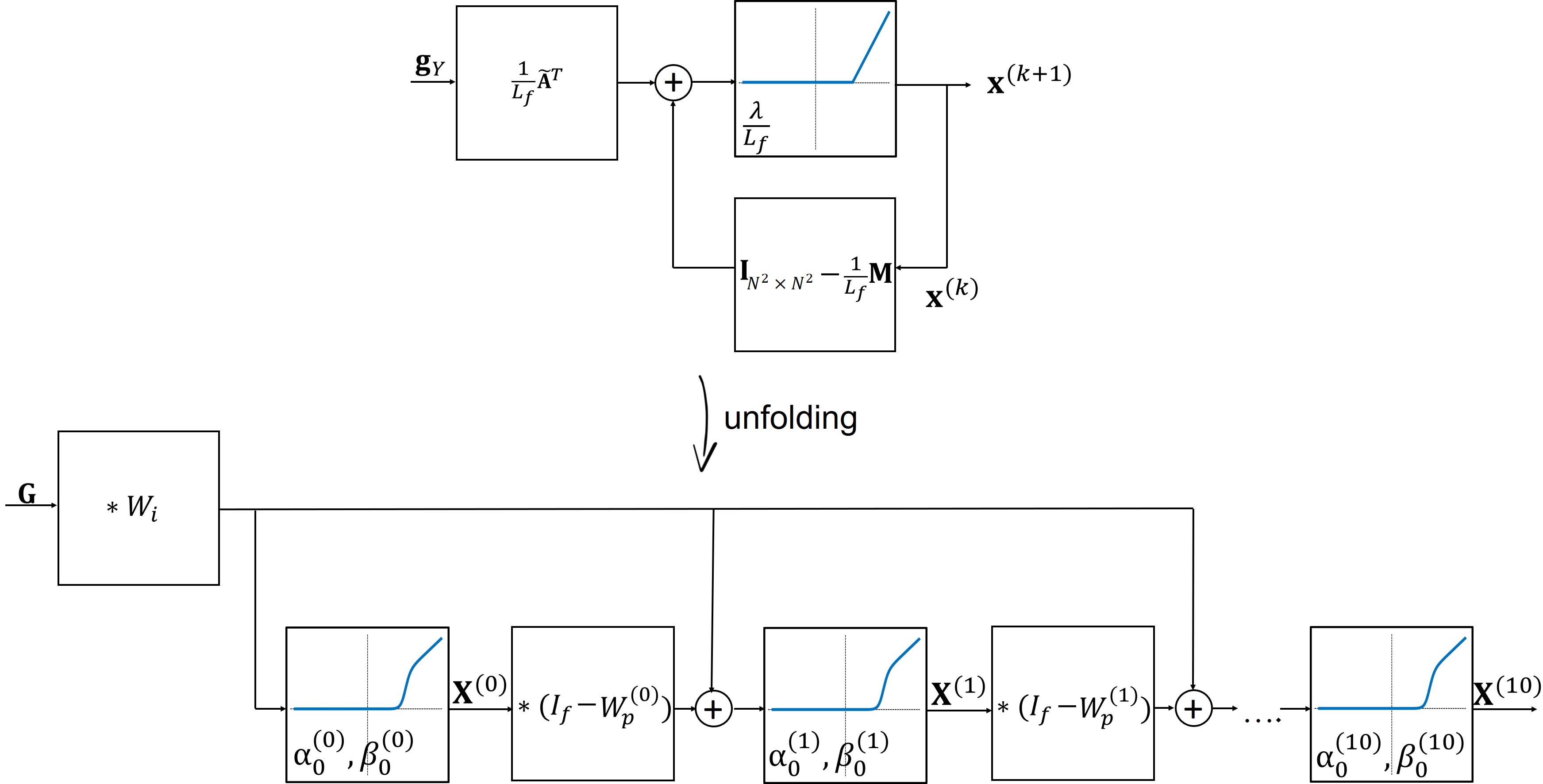} 
\caption{\footnotesize Iterative algorithm vs. unfolded network (LSPARCOM).
Top: block diagram of the SPARCOM algorithm via ISTA, according to Algorithm 1, recovering the vector-stacked super-resolved image $\textbf{x}^{(k)}$. Here, \textbf{v} is calculated according to (15), thus the input $\textbf{g}_Y$ is the temporal variance of the columns of $\textbf{Y}$; the block with the blue graph is $T^+_{\frac{\lambda}{L_f}}(\cdot)$, the positive soft thresholding operator with parameter ${\frac{\lambda}{L_f}}$; the other blocks denote matrix multiplication (from the left side).
Bottom: LSPARCOM, recovering the super-resolved image $\textbf{X}^{(k)}$. The input $\textbf{G}$ is the matrix-shaped resized version of $\textbf{g}_Y$; the blocks with the blue graph apply the smooth activation function $S^+_{\alpha_0,\beta_0}(\cdot)$ with two trainable parameters: $0 \leq \alpha_0^{(k)} \leq1$, $\beta_0^{(k)}$, $k= 0,..,10$, see (17); the other blocks denote convolutional layers, where $I_f$ is a non-trainable identity filter (dirac delta filter) and $W_i$, $W^{(k)}_{p}$, $k= 0,..,9$ are trainable filters.}
\end{figure} 

\begin{algorithm}
\caption{LSPARCOM at inference time}
\begin{algorithmic}
\STATE \textbf{Input:} $\textbf{G}$
\STATE \textbf{Trainable parameters:} $W_i$; $W^{(k)}_p$, $k= 0,..9$; $0 \leq \alpha_0^{(k)} \leq1$, $\beta_0^{(k)}$, $k= 0,..,10$; $s$ 
\STATE \textbf{Initialize:} $\textbf{X}^{(0)} =  S^+_{\alpha_0^{(0)},\beta_0^{(0)}}(\textbf{G}*W_i)$
\STATE \textbf{For} $k = 0:9$
\item $ \textbf{X}^{(k+1)} =  S^+_{\alpha_0^{(k+1)},\beta_0^{(k+1)}}[\textbf{G}*W_i - \textbf{X}^{(k)}*W^{(k)}_p + \textbf{X}^{(k)}]$
\STATE \textbf{end} 
\STATE \textbf{Output:} $\textbf{X}^{(out)} = s\cdot \textbf{X}^{(10)}$ 
\end{algorithmic}
\end{algorithm}

\subsection{Input}

Out of all the inputs to Algorithm 1, $\textbf{v}$ is the only one that contains information regarding the measured data, thus it is a natural input choice to an unfolded deep network that draws inspiration from SPARCOM. However, vector $\textbf{v}$ uses the information regarding the PSF, stored in matrix \textbf{A}, to pick and weight elements of the covariance matrix $\textbf{R}_y$. Therefore, it cannot be used straight-forwardly in our unfolded method, where we assume the PSF is unknown. 
In the degenerate case where the PSF is a dirac delta function in the low-resolution grid, meaning that the presence of an emitter in a certain pixel does not affect its neighboring pixels, (9) approximately selects only the diagonal of the covariance matrix $\textbf{R}_y$. Upon reshaping it as an image, this diagonal is the temporal variance of the input image stack, resized to the high-resolution grid, which we denote as $\textbf{G}$. Thus, in this case $\textbf{v} $ is merely a resized version of the second-order SOFI image with auto-cumulants \cite{SOFI}. This improves upon the resolution of the diffraction-limited image by a factor of $\sqrt{2}$, enabling a better starting point for sparse coding. 
The logic in the choice of off-diagonal elements lies in the fact that assuming the original (low-resolution) pixel size is smaller than a diffraction limited spot, the presence of an emitter in one pixel in the low resolution grid is manifested in the signal obtained in its neighboring pixels. A similar principle is used in the cross-cumulant version of SOFI \cite{SOFI}. 
Thus, assuming the PSF to be a dirac delta function actually means we ignore the effect of adjacent pixels, and observe the time-trace of each pixel separately. While this approach obviously ignores spatial information, it is a reasonable choice for a starting point in case the PSF is unknown.

Seeking for a formulation which does not mandate prior knowledge of the PSF, a similar optimization problem to that given in (8) can be obtained by posing a sparse recovery problem on the variance image directly \cite{SUSHI}. This results from applying the variance operator $\textbf{Var}[\cdot]$ on (2)
\begin{equation} 
\textbf{Var}[\textbf{y}(t)]= \textbf{Var}[\textbf{A}\textbf{s}(t)] = \textbf{A}^2\textbf{Var}[\textbf{s}(t)],
\end{equation}
where each element of \textbf{A} is squared independently (element-wise). Each column of the new measurement matrix $\tilde{\textbf{A}}$ $\overset{\Delta}{=}\textbf{A}^2$
is an $M^2$ long vector corresponding to an $M$ $\times M$ image of the squared PSF shifted to a different location on the high resolution grid, signifying the resolution improvement of the variance image over the original diffraction-limited image by a factor of $\sqrt{2}$ (considering a Gaussian PSF). 

Denoting $\textbf{g}_Y\overset{\Delta}{=}\textbf{Var}[\textbf{y}(t)]$, and using the existing notation for the variance of the emitter fluctuation on a high-resolution grid $\textbf{x} = \text{diag}\{\textbf{R}_s\}=\textbf{Var}[\textbf{s}(t)]$, we obtain:
\begin{equation} 
\textbf{g}_Y= \tilde{\textbf{A}}\textbf{x}.
\end{equation}
Using (13) in place of (6) in the optimization problem (7), we replace the function $f(\textbf{x})$ by
\begin{equation} 
f(\textbf{x}) = \frac{1}{2}\|\textbf{g}_Y - \tilde{\textbf{A}}\textbf{x} \|_F^2,
\end{equation}
yielding the classical compressed sensing ISTA problem \cite{ISTA1, LISTA}. Minimizing (7) with the new expression (14) for $f(\textbf{x})$ can be performed as shown in Algorithm 1, with only the inputs changed. According to the gradient of (14), \textbf{v} becomes
\begin{equation} 
\textbf{v} = \tilde{\textbf{A}}^T\textbf{g}_Y, 
\end{equation}
and $\textbf{M}$ is replaced by
 \begin{equation} 
\textbf{M} = \tilde{\textbf{A}}^T\tilde{\textbf{A}},
\end{equation} 
with $L_f$ being the Lipschitz constant of the gradient of (14), equal to the maximum eigenvalue of $\textbf{M}$.
Here too, $\textbf{M}$ and $\textbf{v}$ are obtained by differentiating (14) with respect to $\textbf{x}$, resulting in $\nabla$$f(\textbf{x}) = \textbf{M}\textbf{x}-\textbf{v}$.

As can be seen from (15), in this formulation $\textbf{v}$ is obtained by multiplying $\tilde{\textbf{A}}^T$ by the vector-stacked $M \times M$ temporal variance of the input image stack. 
In this manner, we start from the default reconstruction we use in the lack of explicit knowledge of the PSF, which is the temporal variance (implicitly choosing the diagonal), and account for the PSF by applying (15) inside the network. This is feasible since the multiplication operation can be easily replaced by applying a learned convolutional layer to the variance of the input image stack, as explained in the following section. 
Thus, following the formulation of $\textbf{v}$ given in (15), we take $\textbf{G}$, the matrix-shaped resized version of $\textbf{g}_Y$, as the input to the network, which enables replacing the multiplication by the unknown $\tilde{\textbf{A}}^T$ with a learned filter. 
For more details, see Fig. 7 in the appendix, which illustrates $\textbf{v}$ in the scenarios discussed above.

\subsection{Convolutional layers}

In the original paper suggesting algorithm unfolding \cite{LISTA}, the authors unfold the iterative algorithm with minimal modifications, leaving the multiplication operators used in the original algorithm intact. 
Recent publications \cite{LISTA_CONV, CORONA} suggest convolutional extensions to the original approach, allowing a dual benefit: exploiting the shared spatial information between neighboring image pixels, and significantly reducing the number of trainable parameters, leading to more effective training. 

As can be seen from the top part of Fig. 1, in our setting, we wish to replace the following two matrix-multiplication operations with convolutions: multiplication by $\tilde{\textbf{A}}^T$ and multiplication by $\textbf{M}$.
In our case, this replacement is natural and requires no structural compromise, as due to their unique structures, multiplication by both $\tilde{\textbf{A}}^T$ and $\textbf{M}$ is equivalent to convolution with an appropriate kernel. 
To understand this equivalence, lets return to (1). Another way to formulate this equation is:
 \begin{equation} 
f[m\Delta_L,n\Delta_L,t] =u[m\Delta_L,n\Delta_L]*\sum_{i=0}^{N-1} \sum_{l=0}^{N-1}\delta[m\Delta_L-i\Delta_H,n\Delta_L-l\Delta_H]s_{il}(t),
\end{equation}
meaning the true image, composed of the emitters localized on the high-resolution grid and imaged on the low-resolution grid, is blurred by convolution with the PSF to obtain the diffraction-limited image. 
We can formulate (17) in matrix form in an alternative form to (2):
\begin{equation} 
\textbf{y}(t)= \textbf{A}_s\textbf{s}_l(t),
\end{equation}
where $\textbf{s}_l(t)$ is the down-sampled version (to the low resolution grid) of $\textbf{s}(t)$, and $\textbf{A}_s$ is a symmetric square $M^2 \times M^2$ convolution matrix in which each column (and row) is an $M^2$ long vector corresponding to an $M$ $\times M$ image of the PSF shifted to a different location on the low-resolution grid. For example, to get the first entry of $\textbf{y}(t)$, we multiply the first row of $\textbf{A}_s$, which is the vector-stacked image of the PSF shifted to the upper left corner, with $\textbf{s}_l(t)$; this is exactly the same as positioning the convolution kernel (the PSF in this case) in the upper left corner of the image-stacked $\textbf{s}_l(t)$.

Now lets return to the matrix formulation in (2), where we multiply the $M^2 \times N^2$ matrix $\textbf{A}$ by the vector stacking of the high resolution image $\textbf{s}(t)$. 
In matrix \textbf{A} each of the $N^2$ columns is an $M^2$ long vector corresponding to an $M$ $\times M$ image of the PSF shifted to a different location on the high-resolution grid, and each of the $M^2$ rows is an $N^2$ long vector corresponding to an $N$ $\times N$ image of the PSF shifted to a different location on the low-resolution grid. 
Even though \textbf{A} is not square, it has a symmetry resembling that of $\textbf{A}_s$. Practically, since the rows of \textbf{A} are of interest, the multiplication operation in (2) is equivalent to convolving the high resolution image with the PSF interpolated to the high-resolution grid, but with strides of $\frac{N}{M}$, creating a down-sampling effect.

Moving on to $\tilde{\textbf{A}}$, it has the same form as \textbf{A} only that the PSF is squared. Similarly, in the $N^2 \times M^2$ $\tilde{\textbf{A}}^T$, each of the $M^2$ columns is an $N^2$ long vector corresponding to an $N$ $\times N$ image of the squared PSF shifted to a different location on the low-resolution grid, and each of the $N^2$ rows is an $M^2$ long vector corresponding to an $M$ $\times M$ image of the squared PSF shifted to a different location on the high-resolution grid. Practically, since the rows of $\tilde{\textbf{A}}^T$ are of interest, multiplying $\tilde{\textbf{A}}^T$ by $\textbf{g}_Y$ as in (15) is equivalent to convolving the low resolution variance image with the squared PSF (in the low-resolution grid), but with sub-pixel strides of $\frac{M}{N}$, creating an up-sampling effect. This operation is approximately equivalent to convolving the variance image interpolated to the high-resolution grid (\textbf{G}) with the squared PSF interpolated to the high-resolution grid, with unit strides. The latter is preferred over the former due to ease of execution. 
 
Finally, we explore the multiplication by $\textbf{M} = \tilde{\textbf{A}}^T\tilde{\textbf{A}}$; $\textbf{M}$ is a symmetric square $N^2 \times N^2$ matrix, where each column (and row) is an $N^2$ long vector corresponding to an $N$ $\times N$ image of a kernel shifted to a different location on the high-resolution grid. This kernel is composed of the sum of the squared PSFs on a high-resolution grid, shifted and weighted according to the low-resolution squared PSF. Practically, multiplying \textbf{M} by the high resolution vector-stacked image \textbf{x} is equivalent to convolving the high resolution image \textbf{X} with the above-mentioned kernel (with unit stride). 

The convolutional layer trainable parameters include the filters $W_i$ and $W^{(k)}_p$, $k= 0,..9$. 
In the illustration shown in Fig. 1, $I_f$ is a non-trainable identity filter; this is merely a way of maintaining structural similarity to SPARCOM, while compactly adding the output of the previous fold $\textbf{X}^{(k)}$, as explained in Algorithm 2. Practically, instead of using a non-trainable identity filter one can simply use a subtraction layer to subtract $\textbf{X}^{(k)}*W^{(k)}_p$ from $\textbf{G}*W_i$ and another addition layer to add $\textbf{X}^{(k)}$ to the result of the latter.

As shown in Fig. 1, The first matrix multiplication, by $\frac{1}{L_f}\tilde{\textbf{A}}^T$, was replaced by a single convolutional layer, whereas the multiplication by $\frac{1}{L_f}\textbf{M}$ was replaced by a layer-dependent convolutional layer, mostly since sharing these weights across the layers may cause gradient explosion and vanishing problems, similar to recurrent neural networks (RNNs) \cite {UNFOLD_REV}.

Since the choice of a starting point can affect the ability of the learning procedure to converge to a global minima, the various trainable kernels were initialized according to their respective tasks.
The filter $W_i$, which imitates multiplication by $\tilde{\textbf{A}}^T$, and should theoretically be the squared PSF of the training data interpolated to the high-resolution grid, was initialized as a $25 \times 25$ Gaussian filter with $\sigma = 1$.
All filters $W^{(k)}_p$, which imitate multiplication by \textbf{M}, and should theoretically be the sum of the squared PSF of the training data on a high-resolution grid, shifted and weighted according to the low-resolution squared PSF, were also initialized by a Gaussian filter with $\sigma = 1$, but with a larger window size of $29 \times 29$.
The exact window size of the filters, determining truncation, was determined by testing the iterative version using the known kernels, and choosing the smallest window size not causing significant quality degradation.

To reduce the number of parameters in the model given the large kernel sizes, we utilized the reasonable assumption of radially symmetric kernels, and added a radial constraint to the kernel weight update procedure.
Practically, we did that by replacing the values of all locations at equal euclidean distance from the center of the kernel by their average value. 
Note that the assumption of radial symmetry may be inaccurate in case of PSF aberrations which could result from imperfect alignment of the optics, yet the ability to accurately describe the PSF under this radial constraint is still superior to that enabled by classical Gaussian-based PSF estimation methods \cite {PSF_ESTIMATION}.
The exact values of the filters, calculated from the PSF of the training data, vs. their learned versions are given in Fig. 8 in the appendix.

\subsection{Activation function}

The positive soft thresholding operator used in Algorithm 1, which is the proximal operator for the $L_1$ norm, is a shifted ReLU, and thus can trivially be incorporated in the neural network as a non linear activation layer. Nevertheless, to increase sparsity we replace it with a differentiable sigmoid-based approximation of the positive hard thresholding operator \cite{HTH_APP}, which is the proximal operator for the $L_0$ norm, thus effectively changing the $L_1$-regularizer $\|\cdot\|_1$ in (7) to an $L_0$-regularizer $\|\cdot\|_0$. This smooth activation function, denoted $S^+_{\alpha,\beta}(\cdot)$, has two trainable parameters ($\alpha,\beta$), differing between folds, allowing both the slope and the cutoff to be learned.

Since different parts of an image (and of course different images) typically have different emitter intensities, e.g. due to inhomogeneous illumination, we defined the trainable values of the smooth activation function to be relative rather than absolute values, noted as $\alpha_0$ and $\beta_0$, where $0 \leq \alpha_0 \leq1$; this allows applying a local rather than global threshold for each processed patch. In this setting, the first and 99th percentile $i_1,i_{99}$ of each channel of the input to the activation layer are calculated, giving an estimation of the largest and smallest values in the respective channel of the patch, excluding noise. Then, these values are used to estimate the local value of the cutoff $\alpha$ for each channel according to:
\begin{equation} 
\alpha = i_1 + (i_{99} - i_1)\alpha_0,
\end{equation}
and the local value of the slope $\beta$ for each channel as
\begin{equation} 
\beta = \frac{\beta_0}{\alpha}.
\end{equation}
For example, if training yields $\alpha_0=0.5$ the cutoff would be the average of the first and 99th percentile in the patch.

The resulting activation function is then given as a function of the local $\alpha$ and $\beta$:
\begin{equation} 
S^+_{\alpha,\beta}(x) \overset{\Delta}{=} \frac{ReLU(x)}{1+exp[-\beta(|x|-\alpha)]}.
\end{equation}
In (21) $x$ is scalar; when applied to vectors and matrices $S^+_{\alpha,\beta}(\cdot)$ operates element-wise.
Figure 9 in the appendix shows $S^+_{\alpha=5,\beta}(x)$ for various values of the slope $\beta$, relative to a shifted ReLU (positive soft threshold) and a positive hard threshold.

\subsection{Learning process}

LSPARCOM has multiple trainable parameters (see Algorithm 2): $W_i$; $W^{(k)}_p$, $k= 0,..9$; $0 \leq \alpha_0^{(k)} \leq1$, $\beta_0^{(k)}$, $k= 0,..,10$, $s$. These parameters are learned by feeding the network a set of inputs along with their ground truth outputs, i.e. a training set, which can generally be obtained either by applying other methods or by using simulated data. 

Learning the trainable parameters in LSPARCOM is done using the Adam optimization algorithm \cite{ADAM}, which is an extension of the well-known stochastic gradient descent algorithm. This supervised learning procedure is composed of two parts: a forward pass and a backward pass. 

The forward pass begins by using the initial values of the trainable parameters: for the convolutional filters these values are explained in Section 4.2; for the activation layers we initialized $\alpha_0 = 0.95$ and $\beta_0 = 8$ for all layers; and for the final scaling layer we initialized $s=0.01$. 
The forward pass ends by measuring the quality of the reconstruction $\textbf{X}^{(out)}$ by comparing it to the corresponding ground truth image $\textbf{X}^{(GT)}$ via a loss function. 
For that purpose, we define a binary mask $\textbf{B}$, created by binarizing $\textbf{X}^{(GT)}$. Then, the loss for the supervised weight optimization procedure is defined as a function of the prediction $\textbf{X}^{(out)}$ and the ground truth $\textbf{X}^{(GT)}$ as follows: 
\begin{equation} 
Loss = 
\frac{1}{N^2}\sum_{i,j = 1}^{N} \textbf{B}(i,j)[\textbf{X}^{(GT)}(i,j) - \textbf{X}^{(out)}(i,j)]^2 + \lambda [1-\textbf{B}(i,j)]|\textbf{X}^{(out)}(i,j)|,
\end{equation}
in similar spirit to the unsupervised optimization problem defined in (7). 
As a result of using the mask $\textbf{B}$, the first element is zeroed for pixels that do not contain emitters, and the second is zeroed for pixels that do. 
The regularization factor $\lambda$ balances between the demand for accurate reconstruction in pixels that contain emitters, and the demand for minimal false detections; its value depends on the magnitude of the data. The choice of $\lambda$ only has to be made once, for the training set; as demonstrated in the following section, the network generalizes well to different data types, using the exact same weights that were optimized for a loss using a specific value of $\lambda$ suitable for the training set, even when the $\lambda$ parameter used in SPARCOM differs significantly. 

The backward pass computes the gradient of the loss function with respect to the trainable parameters, based on applying the chain rule for derivatives \cite{DL0}; this allows updating the trainable parameters in a manner that minimizes the loss.
The updated parameters are then used to do another forward pass to calculate the new loss, another backward pass to update the weights based on this loss, and so on, until convergence.

We trained the weights with the Adam optimization algorithm \cite{ADAM} with $\beta_1=0.9$, $\beta_2=0.999$, and an initial learning rate of $1e^-4$, using the Keras library running on top of TensorFlow.

\section{Results}

In this section, we compare the reconstruction quality achieved by using LSPARCOM to that achieved by the iterative algorithm it is based on, SPARCOM \cite{SPARCOM}, as well as to the leading classical deep-learning based localization method for high emitter-density frames, Deep-STORM \cite{DEEP-STORM}. 
All algorithms tested for comparison use high emitter-density frames as inputs, and thus allow similarly high temporal resolution; thus, the comparison focuses on the quality of the reconstruction based on the exact same input frames. 
Each of the three techniques compared require different prior knowledge as additional input, the details of which are explained in Section 2.1. 
When possible, we also compare the results to the ground truth localization.

We further compare runtimes for all algorithms; for that purpose, we implemented all algorithms serially in Matlab. All timings were conducted while running on a 32GB Intel Core i7-4790 CPU @ 3.6 GHz, and can be accelerated by running on a designated environment inducing parallelism using a GPU.

For figures 2,4,5 and 6, the colormap is such that white corresponds to the highest value, then yellow, red and black. For the SPARCOM and LSPARCOM reconstructions, the value obtained corresponds to the variance of the emitter, while for the ground truth, Deep-STORM and ThunderSTORM reconstructions, the value obtained corresponds to the integrated emitter intensity. Since the value itself is usually not of interest, but rather only the support of the matrix, indicating the location of the emitters, the maximal and minimal values mapped to the edges of the colormap at each image were chosen to obtain optimal visibility.

\subsection{Training, pre-processing and prior knowledge}

\subsubsection {LSPARCOM}
As explained in Section 4, in LSPARCOM the super-resolved image is generated by inputting a single image, which is constructed by calculating the temporal variance of all the high density frames. 

We trained two different networks.
Each of the two training sets we used included overlapping patches taken from multiple frames composing a single FoV with an underlying structure. 
The use of a single FoV for training was made possible by the use of small patches. 

The image used for training the first network was a synthetic dataset simulating biological microtubules \cite{EPFL2015}, which is publicly available. 
The ground truth positions and diffraction limited image can be seen in Fig. 2 (the diffraction limited image in Fig. 2(a)); we refer to this training set as BT (bundled tubes).
The diffraction limited image stack we used included 12,000 $64 \times 64$ pixel low-density frames, with a density of 0.16 emitters per $\mu m^2$ (0.0016 emitters per square low-resolution pixel), representing a single FoV, with coinciding emitter locations. We randomly summed them to get various combinations of 40 $\times$ higher density frames, to obtain 360 high-density frames per input ($\sim6.6$ emitters per $\mu m^2$). 

The image used for training the second network was a synthetic dataset simulating biological tubulins, generated with the exact same imaging parameters as the ones Deep-STORM net2 was trained on (see 5.1.3), using the ThunderSTORM ImageJ plugin \cite {ThunderSTORM}, based on the structure publicly available in \cite{EPFL2015}.
The ground truth positions and diffraction limited image are based on the ones shown in Fig. 4, yet were generated by summing 4 such identical FoVs, rotated by either 0, 90, 180 or 270 degrees, to create a more dense FoV (see Fig. 10 in the appendix); we refer to this training set as TU (tubulins).
The diffraction limited image stack we used included 2,000 $384 \times 384$ pixel high-density frames, with a density of 1.4 emitters per $\mu m^2$ (0.014 emitters per square low-resolution pixel), representing a single FoV, with coinciding emitter locations. We randomly summed them to get various combinations of 5 $\times$ higher density frames, to obtain 350 very high-density frames per input ($\sim7$ emitters per $\mu m^2$). 

In this work, we used an upsampling factor of $\frac{N}{M}=4$. 
To create training diversity, each input was randomly rotated by either 0, 90, 180 or 270 degrees, then the variance was calculated both for the input and ground truth stack (and resized for the input), and a random $64 \times 64$ pixel window, corresponding to a $16 \times 16$ window in the diffraction-limited image stack, was cropped. 
We trained both networks on relatively small $M=16$ patches, with the rationale that distant parts of the image, which are further away than the width of the PSF, should not effect one-another, and more diverse and effective training can be obtained using small patches.
A total number of 10000 such patch-stacks, each corresponding to a $16 \times 16$ window in the high-density diffraction-limited image stack were used for training. 

The patch approach was also used for testing, where patch recombination was improved by using overlapping patches and weighting with a Tukey (tapered cosine) window. 

The 10 folds of LSPARCOM used here include overall 9058 trainable parameters, which were reduced to only 1166 parameters given the radial constraint. For example, a $29 \times 29$ kernel, which usually consists of $29^2 = 841$ parameters only included 106 parameters under the radial constraint, imposing all locations in equal euclidean distance from the center pixel to be equal.

LSPARCOM was trained twice, each time from a different single FoV, as explained above, yielding two sets of weights that were used for all tests. 
We trained both versions over 750 epochs, each of which took approximately 12 hours on a PC equipped with an NVIDIA GeForce GTX1050 GPU and a 32GB Intel Core i7-4790 CPU @ 3.6 GHz. 
For the BT LSPARCOM net we used a regularization parameter of $\lambda= 0.7$, and for the TU LSPARCOM net - $\lambda= 1$. 

Pre-processing of the raw data, both prior to training and testing, includes 2 simple steps: normalizing the movie intensity to have a maximal value of 256, and removing the temporal median of the movie from each frame.

\subsubsection {SPARCOM}

For all tests, the classical version of SPARCOM was executed over 100 iterations \cite{SPARCOM}, without using a weighing matrix or applying the sparsity prior in another transformation domain \cite{SPARCOM2}, to keep the basic algorithm similar to the one used for unfolding. 
To achieve optimal performance we used the original Algorithm from \cite{SPARCOM}, using the Fourier-domain implementation and the fast ISTA (FISTA) \cite{FISTA, FISTA2}.

Similar to LSPARCOM, we used a patch approach (for computational reasons), where patch recombination was improved by using overlapping patches and weighting with a Tukey (tapered cosine) window. 
We followed the code released by the authors \cite{SPARCOM-CODE}, using zero padding, which induced a slightly higher number of patches than that used in LSPARCOM (where no padding was needed).

The regularization factor $\lambda$ was hand-picked and fine-tuned to fit each tested data type.
For each test set we used two versions of SPARCOM: one where we assumed the PSF is unknown and inputted a dirac delta function instead, and the other using explicit prior knowledge of the PSF. In the latter case, for the simulated data the PSF used for generating the data was accurately given; for the experimental data we used the same PSF as the simulated data, assuming it is similar. 
Note that for experimental data of high emitter-density frames good approximation of the PSF may be impossible. Nevertheless, it can usually be estimated as a Gaussian filter with a standard deviation calculated based on the numerical aperture of the objective lens and the fluorophore emission wavelength, if these are known \cite{PSF_ESTIMATION}. Our choice of presenting a version of SPARCOM where the PSF is estimated as a dirac delta function aims to demonstrate the improvement obtained by LSPARCOM over simply selecting the diagonal of the covariance matrix, as explained in Section 4.1 discussing the input. 

Similar to LSPARCOM, pre-processing in SPARCOM includes normalizing the movie intensity to have a maximal value of 256, and removing the temporal median of the movie from each frame.

\subsubsection {Deep-STORM}

In Deep-STORM \cite{DEEP-STORM} the final super-resolved image is created by summing a set of independently reconstructed high-density frames with localized emitters. The training set includes a few thousands patches of randomly distributed emitters generated via the ThunderSTORM ImageJ plugin \cite{ThunderSTORM}, ideally with the exact same imaging parameters as the test set. 
These parameters include the camera base level, photoelectrons per A/D count, PSF, emitter FWHM range, emitter intensity range and mean photon background.

The Deep-STORM net architecture is based on a fully convolutional encoder–decoder network, and includes 1.3M trainable parameters, used for the convolutional and batch-normalization layers. 
Two versions of Deep-STORM, with weights trained according to different imaging parameters, were used for testing; for future reference, we refer to them as Deep-STORM net1 and Deep-STORM net2. 
To assure optimal training for a fair comparison, we used the two trained versions released in the demo by the authors \cite {DEEP-STORM-CODE}. 

Pre-processing in Deep-STORM includes resizing the input frame to the desired dimensions of the output (determining the final grid size), projecting each frame to the range $[0$ $1]$, and normalizing each frame by removing the mean value of the training dataset and dividing by its standard deviation, as specified in the code released by the authors \cite {DEEP-STORM-CODE}.

\subsection{Simulation results}

In this subsection we compare the methods on realistic data generated by simulation, such that the ground truth localizations are known. The exact knowledge of the ground truth allows us to quantify the quality of the results using the signal-to-noise ratio (SNR) metric, which was also used in the SMLM challenge \cite{EPFL2015}. The SNR is defined as follows:
\begin{equation} 
\text{SNR} \overset{\Delta}{=} 10\text{log}_{10}\frac{\|\textbf{X}^{(GT)}\|^2}{\|\textbf{X}^{(GT)}-\textbf{X}^{(out)}\|^2}
\end{equation}
where the norm is calculated on the vectorized matrix.
Since this metric compares intensities and not just emitter locations, whereas in our case the exact intensity value is a byproduct of the method of choice, meant to signify emitter locations, and is not important on its own, we binarized all images prior to calculating the SNR; the thresholding step was optimized to obtain the best possible score per each method and dataset.

The inputs to the different algorithms, which were used to generate the figures in this section, as well as the corresponding diffraction limited images, are shown in Fig. 11(a-d) in the appendix. 

\begin{figure}[h] 
\centering
\includegraphics[scale=0.37]{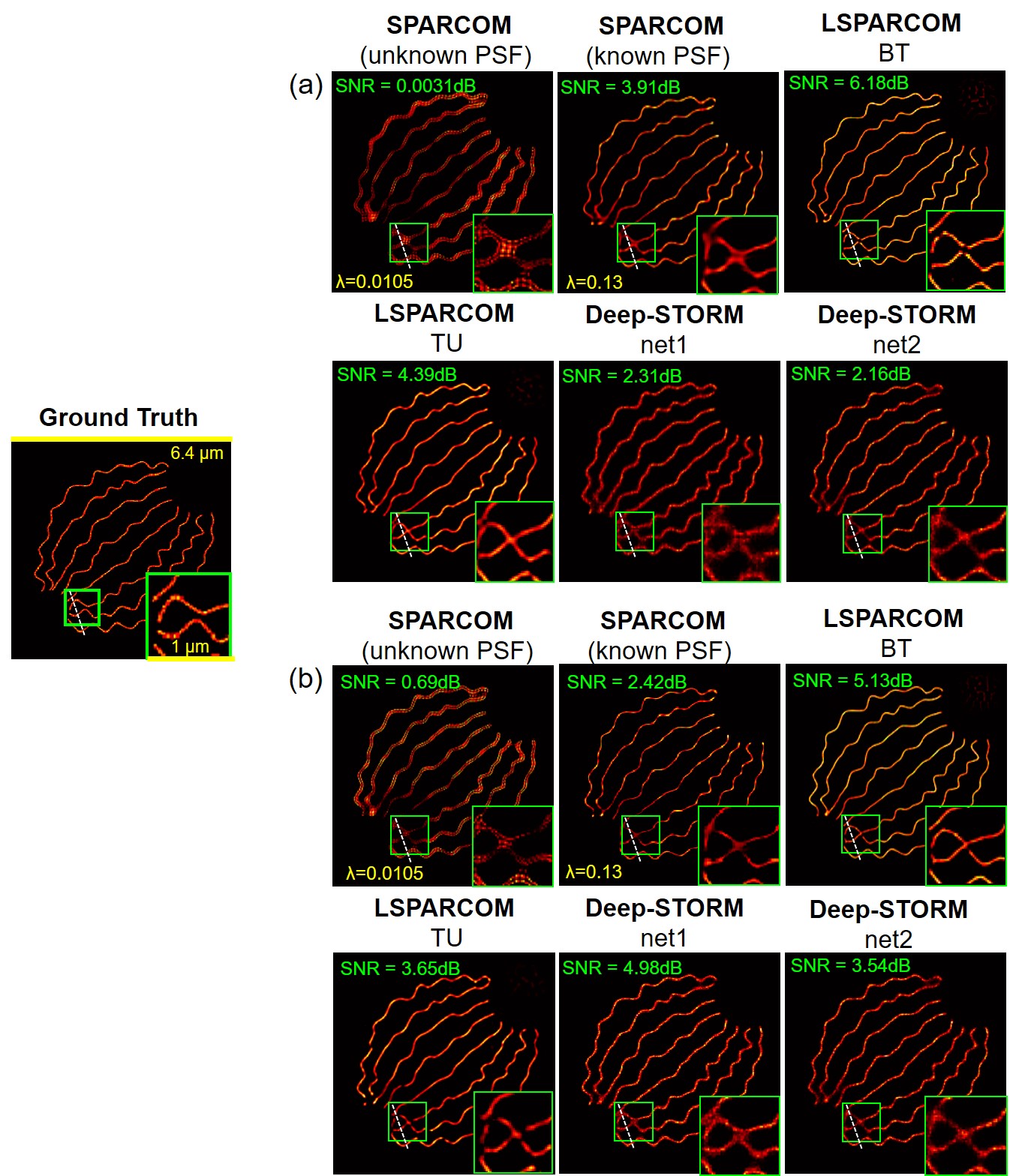}
\caption{\footnotesize Performance evaluation for simulated biological microtubules dataset, composed of 361 high-density frames. The most difficult area for reconstruction is shown magnified in the green box. 
Left panel - Ground truth positions. 
Right panel -  (a): imaging parameters used for training the LSPARCOM BT net, (b): imaging parameters used for training Deep-STORM net1.
The SPARCOM reconstruction with unknown PSF (assuming a dirac delta PSF) was executed over 100 iterations with $\lambda = 0.0105$. 
The SPARCOM reconstruction using the correct PSF was executed over 100 iterations with $\lambda = 0.13$.
The dotted white lines corresponds to the intensity section shown in Fig. 3. }
\end{figure} 

Figure 2 shows the results for a structure identical to that used for creating the patches for training LSPARCOM BT (biological microtubules  \cite{EPFL2015}), with two slightly different sets of imaging parameters (simulating imaging with two different microscopes).
Each of the datasets, corresponding to a different set of imaging parameters, is composed of 361 high-density frames.
Fig. 2(a) shows the the first set of imaging parameters, which matches exactly the ones that the LSPARCOM BT net was trained on; whereas Fig. 2(b) shows the the second set of imaging parameters, which matches exactly the ones that Deep-STORM net1 was trained on. The latter dataset was generated based on the structure of the former using the ThunderSTORM ImageJ plugin \cite {ThunderSTORM}.
As can be seen, the performance of Deep-STORM net1 is significantly degraded due to the slight difference in imaging parameters, which is not even visible to the naked eye, whereas the LSPARCOM reconstruction is more robust, yielding good results in both cases. Naturally, LSPARCOM BT, which was trained on similar data, yields better results than LSPARCOM TU for both sets of imaging parameters, yet LSPARCOM TU performs well overall, despite the significantly different training data. Note that for LSPARCOM TU slight fragmentation is visible in the enlarged box for the second set of imaging parameters.
Both LSPARCOM reconstructions do exhibit some false positive localizations in the upper right corner, which is the result of an empty patch and the relative threshold parameters used in the activation layer.
The Deep-STORM net2 reconstruction, which was trained on the same imaging parameters as LSPARCOM TU, is unable to properly separate close microtubules. 
The SPARCOM reconstruction without prior knowledge of the PSF is grainy and inaccurate, but is dramatically improved upon using the correct PSF.

\begin{figure}[h]  
\centering
\includegraphics[scale=0.4]{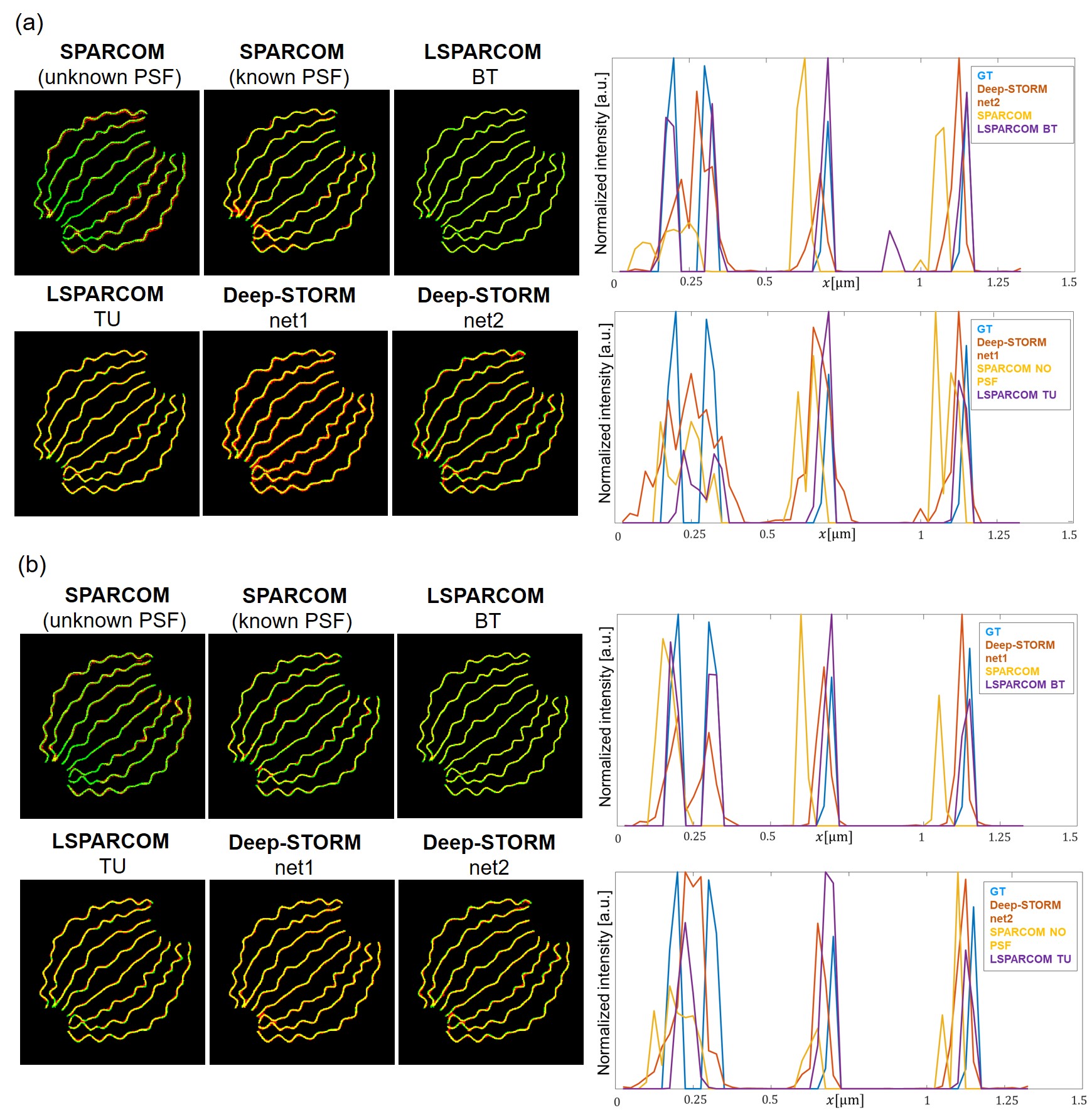}
\caption{\footnotesize Left: Reconstruction from Fig. 2 overlaid with ground truth positions. 
Red: reconstruction; Green: Ground truth; Yellow areas of similar intensity.
Right: Normalized intensity along the dotted white lines in Fig. 2; each comparison is divided to best (top panel) and worst (bottom panel) performers.
Here, (a) corresponds to Fig. 2(a), and (b) corresponds to Fig. 2(b).}
\end{figure} 

The left part of Fig. 3 shows the reconstruction from Fig. 2 overlaid with ground truth positions, where the reconstruction is shown in red, the ground truth localization is shown in green, and areas of similar intensity are shown in yellow.
Prior to the comparison, all reconstructions were binarized using appropriate thresholds, enabling disregarding false positive detections with very low intensity values.
As can be seen, the only reconstruction that allows accurate localization is LSPARCOM BT, with virtually no red zones. While this is to be expected due to the similarity between the training and test set, the next top performer is LSPARCOM TU, in spite of its inherently different training data. For Fig. 2(a) LSPARCOM TU is significantly better than all other methods, whereas in Fig. 2(b) both versions of Deep-STORM come close to its performance. 
Nevertheless, the microtubule reconstruction obtained by LSPARCOM TU induces a slight straightening affect, due to the significantly different training data, under-evaluating the amplitude of the ``curls''. This occurs in most reconstructions, excluding LSPARCOM BT.  
The right part of Fig. 3 shows the normalized intensity cross section along the white dotted lines in Fig. 2, further validating the success of LSPARCOM over all other methods. For visibility, we compare the top performers of each method to the ground truth (GT) in a different panel than the worst performers. 
For Fig. 2(a), the top performers from each method were SPARCOM with explicit knowledge of the PSF, LSPARCOM trained on microtubules (BT), and Deep-STORM net2. Their comparison to the ground truth is shown in the top right panel of Fig. 3(a). As can be seen, LSPARCOM BT achieves optimal localization, coinciding with the peaks of the GT; there is only one outlier, corresponding to a low-intensity false detection. 
For the worst performers of each method, shown in the bottom right panel of Fig. 3(a), LSPARCOM TU clearly achieves the best localization.
For Fig. 2(b), the top performers from each method were SPARCOM with explicit knowledge of the PSF, LSPARCOM trained on microtubules (BT), and Deep-STORM net1. Their comparison to the ground truth is shown in the top right panel of Fig. 3(b). As can be seen, in this case as well LSPARCOM BT achieves optimal localization, coinciding with the peaks of the GT, with Deep-STORM net1 being a close second. 
For the worst performers of each method, shown in the bottom right panel of Fig. 3(b), LSPARCOM TU still achieves the best localization, in spite of the fact that it misses the second peak from the left.
In terms of runtime, the SPARCOM reconstruction took 9.56 sec, the LSPARCOM reconstruction took 2.26 sec, and the Deep-STORM reconstruction took 49.25 sec for $361$ $64\times64$ input frames. 

Figure 4(a) shows the results for a simulated biological tubulin dataset, composed of 350 high-density frames, generated with the exact same imaging parameters as the ones Deep-STORM net2 and LSPARCOM TU net were trained on, using the ThunderSTORM ImageJ plugin \cite {ThunderSTORM}, based on the structure publicly available in \cite{EPFL2015}. 
The top performer in this test is clearly the LSPARCOM TU net, yet this is to be expected due to its training set being based on a similar underlying structure (the test set summed and rotated) with identical imaging parameters.
Even though Deep-STORM net2 was also trained for the exact imaging parameters used and LSPARCOM BT used no prior knowledge, and was trained on an inherently different underlying structure (microtubules), the reconstruction obtained by LSPARCOM BT is significantly more continuous, smooth, and similar to the actual structure. Nevertheless, the LSPARCOM BT reconstruction does suffer from two artifacts: first, it has a slight halo effect; and second, the intersection areas are reconstructed as small arcs, a result of the inherently different training data. 
The reconstructions obtained by the optimal versions of SPARCOM (with knowledge of the PSF) and Deep-STORM (net2, trained on the same imaging parameters) are also good, but more sparse than the ground truth.
The reconstructions obtained by the inferior versions of SPARCOM (without knowledge of the PSF) and Deep-STORM (net2, trained on different imaging parameters) are significantly worse, exhibiting thickened tubulins.
Figure 4(b) presents the reconstruction for a $14 \times$ denser stack, obtained by summing groups of 14 frames in the initial stack, yielding an overall of 25 highly dense input frames. 
While this extremely dense input causes degradation in some of the methods, both LSPARCOM reconstruction remain excellent, where the LSPARCOM BT reconstruction is actually improved, with the arcing affect gone. Deep-STORM net2 also continues to yield a good result. 
In terms of runtime,  the SPARCOM reconstruction took 340 sec, the LSPARCOM reconstruction took 92 sec, and the Deep-STORM reconstruction took 1788 sec for $350$ $384\times 384$ input frames. 
The reconstruction for the $14 \times$ denser stack, including only 25 frames $384\times 384$ input frames took significantly less time with Deep-STORM, since it operates on each frame individually, and took only 125 sec; the runtimes for SPARCOM and LSPARCOM remained nearly constant for the dense stack, as the number of input frames does not influence the runtime.

\begin{figure}[h] 
\centering
\includegraphics[scale=0.43]{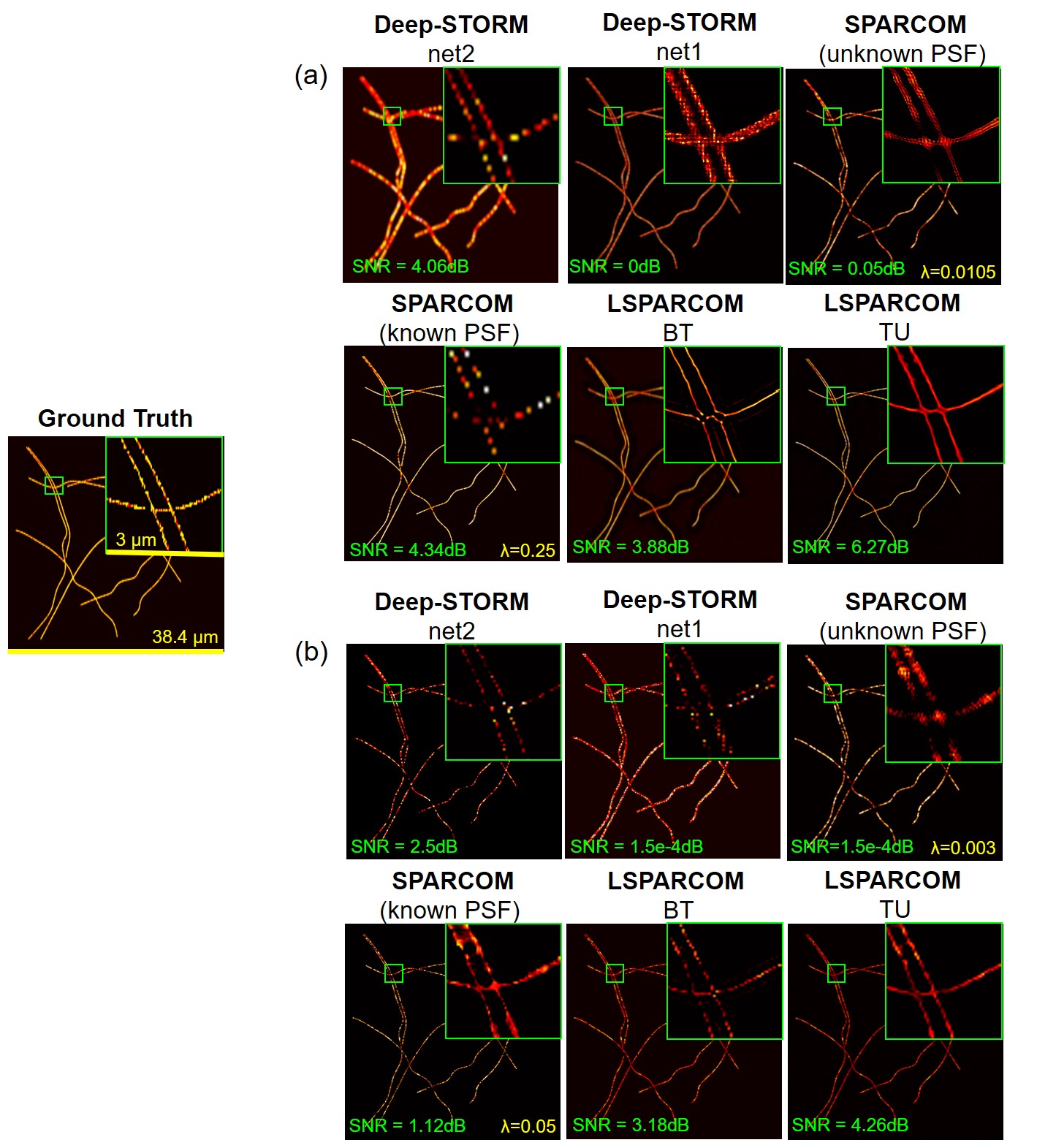}
\caption{\footnotesize Performance evaluation for simulated biological tubulins dataset, composed of (a) 350 high-density or (b) 25 very high-density frames, with the exact same imaging parameters LSPARCOM TU and Deep-STORM net2 were trained on. 
SPARCOM reconstruction with unknown PSF (assuming a dirac delta PSF) was executed over 100 iterations with $\lambda = 0.0105$ for 350 frames, and $\lambda = 0.003$ for 25 frames. 
SPARCOM reconstruction using the correct PSF was executed over 100 iterations with $\lambda = 0.25$ for 350 frames, and $\lambda = 0.05$ for 25 frames. 
LSPARCOM BT net was trained on the dataset from Fig. 2(a), whereas LSPARCOM TU net was trained on a dataset created by summing 4 rotated versions of the FoV from this figure, as in Fig. 10 (with the same imaging parameters).
Deep-STORM net2 was trained on the exact same imaging parameters as the test set, whereas Deep-STORM net1 was trained on different imaging parameters than the test set.}
\end{figure} 

In summary, the simulations shown demonstrate that LSPARCOM yields a robust and accurate reconstruction, comparable or better than the top performers, but with none of the prior knowledge regarding the PSF or imaging parameters, at a fraction of the runtime. Amazingly, LSPARCOM is able to yield excellent reconstruction with as few as 25 high-density frames. The fact that the same version of LSPARCOM (trained from a single FoV) was used to achieve excellent results across different biological structures with variable densities suggests that dedicated training per structure is not required (even though it is able to boost performance).

\subsection{Experimental results}

We next compare all methods on experimental data that is publicly available \cite{EPFL2015}.
For high-density input frames (Fig. 5), there is no ground truth image available for comparison; however, for low-density input frames (Fig. 6) we used the original, low-density frames to create an approximated ground-truth image, using the ThunderSTORM ImageJ plugin \cite{ThunderSTORM}, prior to summing them to get high-density input frames.

The inputs to the different algorithms, which were used to generate the figures in this section, as well as the corresponding diffraction limited images, are shown in Fig. 11(e-h) in the appendix. 

Figure 5 presents the results for an experimental tubulin sequence, composed of 500 high-density frames \cite{EPFL2015}, which was imaged with similar imaging parameters to the ones Deep-STORM net2 and LSPARCOM TU were trained on; more specifically, these imaging parameters correspond to the reported acquisition parameters (such as CCD pixel size) and the observed density, signal and background values. 
While the best results seem to be obtained by the versions of Deep-STORM, SPARCOM, and LSPARCOM using prior knowledge (Deep-STORM net2 and LSPARCOM TU which were trained for similar imaging parameters and SPARCOM with known PSF), LSPARCOM BT yields similar results with no prior knowledge regarding the PSF or other imaging parameters, and no need for heuristic parameter selection. Nevertheless, a slight arcing affect is visible, as in Fig. 4, a result of the inherently different training data. 
Deep-STORM net1 and the version of SPARCOM not using the PSF both yield significantly more blurry, less sparse results, with SPARCOM being the worst of the two.
In terms of runtime, SPARCOM reconstruction took 39.32 sec, LSPARCOM reconstruction took 10.8 sec, and Deep-STORM took 280.38 sec for $500$  $128\times128$ input frames. 

\begin{figure}[h] 
\centering
\includegraphics[scale=0.45]{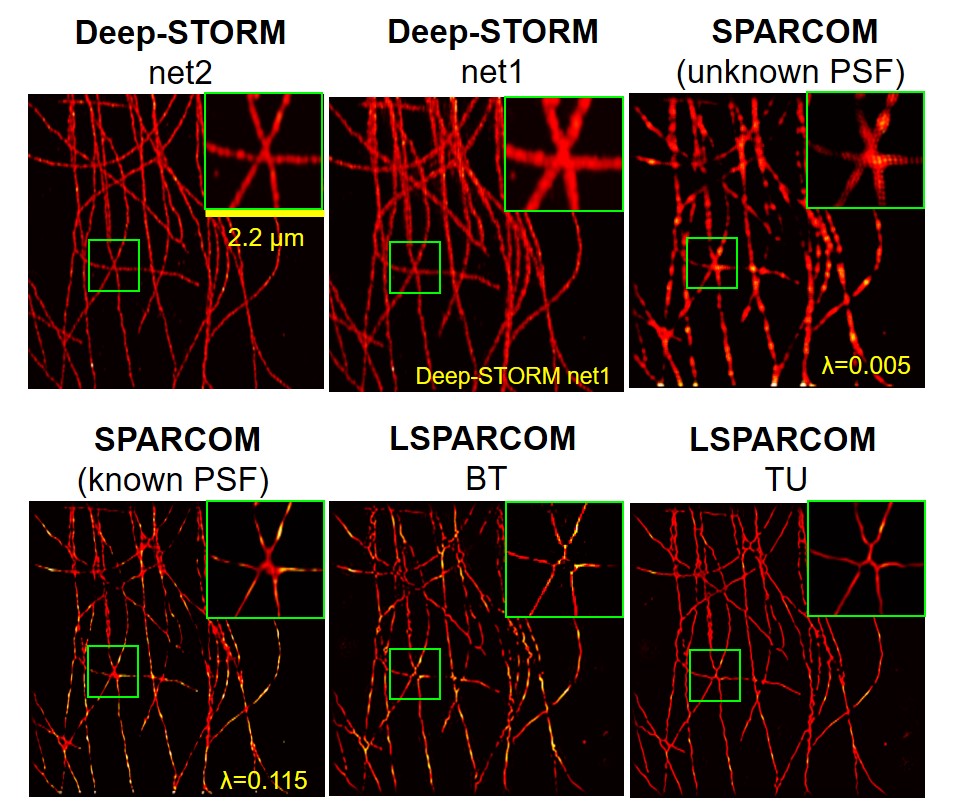}
\caption{\footnotesize 
Performance evaluation for experimental tubulin sequence, composed of 500 high-density frames \cite{EPFL2015} with similar imaging parameters to those Deep-STORM net2 and LSPARCOM TU were trained on. A difficult area for reconstruction is shown magnified in the green box.
The SPARCOM reconstruction with unknown PSF (assuming a dirac delta PSF) was executed over 100 iterations with  $\lambda = 0.005$.
SPARCOM using explicit knowledge of the PSF was executed over 100 iterations with $\lambda = 0.115$. 
The LSPARCOM BT net was trained on the dataset from Fig. 2(a), whereas the LSPARCOM TU net was trained on a dataset created by summing 4 rotated versions of the FoV from Fig. 4, as in Fig. 10 (with the same imaging parameters).
Deep-STORM net2 was trained on the imaging parameters evaluated from the test set, whereas Deep-STORM net1 was trained on different imaging parameters than the test set.
}
\end{figure}

\begin{figure}[h] 
\centering
\includegraphics[scale=0.32]{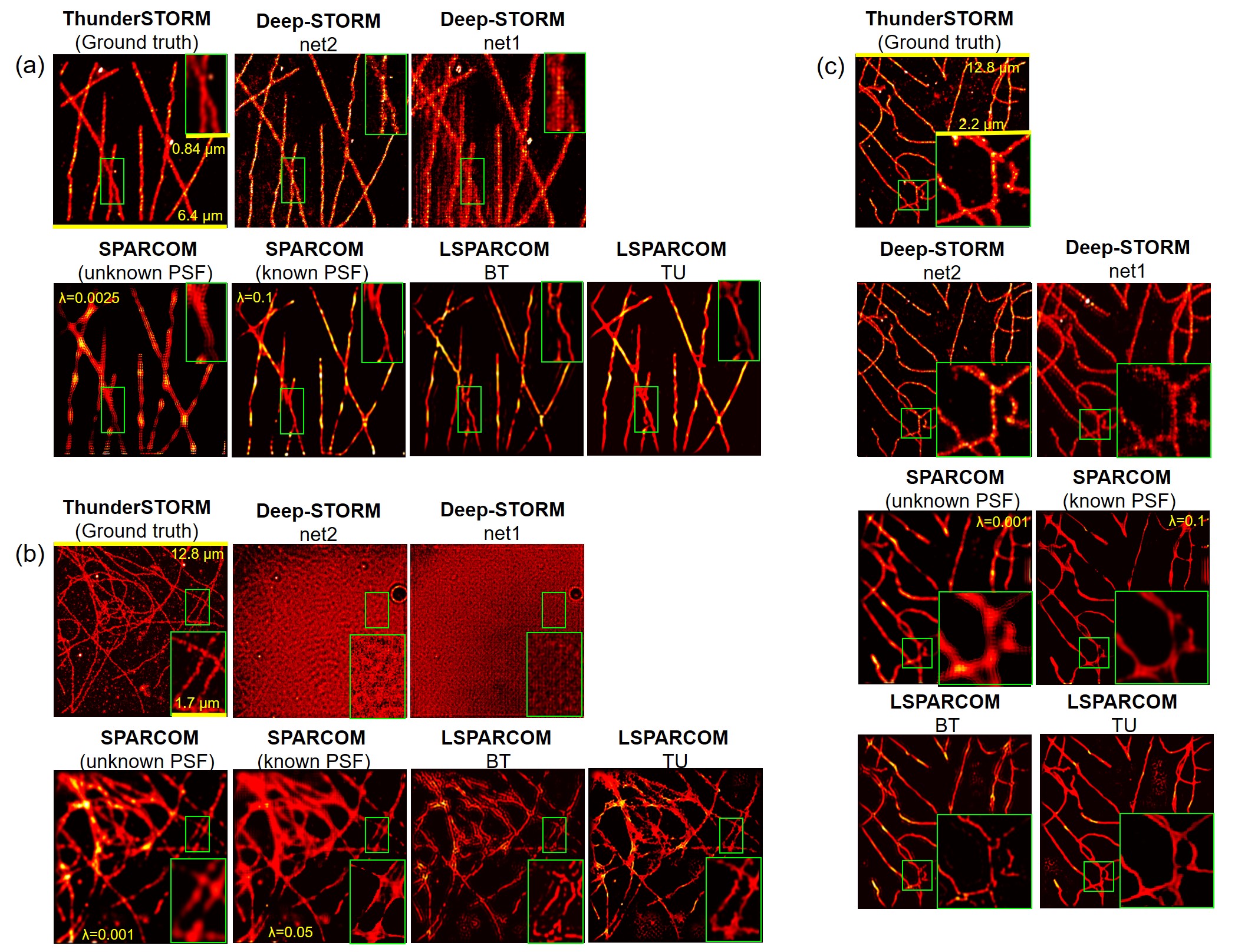}
\caption{\footnotesize 
Performance evaluation for various low-density experimental tubulin sequences \cite{EPFL2015}, summed to get high-density input frames.
(a) 300 high-density frames with similar imaging parameters to those Deep-STORM net2 was trained on. 
(b) 198 high density frames with unknown imaging parameters.
(c) 550 high density frames with unknown imaging parameters. 
The ThunderSTORM reconstruction was obtained from the original low-density frames, and acts as ground truth. 
The SPARCOM reconstruction with unknown PSF (assuming a dirac delta PSF) was executed over 100 iterations with  (a) $\lambda = 0.0025$, (b) $\lambda = 0.001$, and (c) $\lambda = 0.001$.
SPARCOM using explicit knowledge of the PSF was executed over 100 iterations with (a) $\lambda = 0.1$, (b) $\lambda = 0.05$, and (c) $\lambda =  0.1$.
The LSPARCOM BT net was trained on the dataset from Fig. 2(a), whereas the LSPARCOM TU net was trained on a dataset created by summing 4 rotated versions of the FoV from Fig. 4 (as in Fig. 10).
In (a), (c) Deep-STORM net2 and LSPARCOM TU were trained on similar imaging parameters to the test set, whereas in (b) they were trained on different imaging parameters than the test set.
Deep-STORM net1 and LSPARCOM BT were trained on different imaging parameters than the test set for (a)-(c).
}
\end{figure}

Figure 6(a) presents the results for another experimental tubulin sequence, composed of a long sequence of 15,000 low-density frames \cite{EPFL2015}.
The original, low-density frames were used to create a ground-truth image, using the ThunderSTORM ImageJ plugin \cite{ThunderSTORM}. To create high-density frames for testing the performance of the methods in high temporal resolution, we summed each group of 50 frames to obtain 300 high-density frames. 
The reported imaging parameters of this dataset are identical to that shown in Fig. 5, yet many parameters are unreported; nevertheless, they seem to be similar to the dataset Deep-STORM net2 and LSPARCOM TU were trained on. 
As can be seen, both versions of LSPARCOM and the version of SPARCOM that exploits prior knowledge of the PSF and requires fine-tuning the $\lambda$ parameter yield similar good results, yet are in disagreement with the Deep-STORM and ThunderSTORM reconstructions regarding a specific intersection area (see green frame). 
LSPARCOM BT does suffer from three very subtle artifacts: slight halo effect, arcing and fragmentation.
The Deep-STORM net2 gives an overall good reconstruction but a less sparse one, with a higher level of noise arising from false positive localizations. 
Deep-STORM net1, on the other hand, trained on significantly different imaging parameters, yields a blurry, pixelated reconstruction, slightly worse than that obtained by SPARCOM assuming a dirac delta PSF. 
In terms of runtime, SPARCOM reconstruction took 9.68 sec, LSPARCOM took 2.45 sec, and Deep-STORM ran for 49.5 sec for $300$  $64\times64$ input frames. 

Figure 6(b) presents the results for another experimental sequence, composed of a sequence of 9,900 low-density frames. This dataset represents a fixed cell, stained with mouse anti-alpha-tubulin primary antibody and Alexa647 secondary antibody. The intermittent increase in signal is due to reactivation with a 405nm laser \cite{EPFL2015}. The original, low-density frames were used to create a ground-truth image, using the ThunderSTORM ImageJ plugin \cite{ThunderSTORM}. To create high-density frames for testing the performance of the methods in high temporal resolution, we summed each group of 50 frames to obtain 198 high-density frames. 
The reported imaging parameters of this dataset are identical to that shown in Figs. 5 and 6(a); yet most parameters are unreported and remain unknown. 
As can be seen, SPARCOM yields good results only in about half the FoV; this is the case both with and without explicit knowledge of the PSF, suggesting that the PSF used for reconstruction may be inaccurate, a realistic obstacle of experimental data. These reconstructions are similar to the one obtained by LSPARCOM TU, which also suffers from many false negative detections in the bottom left corner. 
LSPARCOM BT yields overall good results, but with a ``chaining'' artifact -  some lines are reconstructed as chains; this artifact is characteristic of having test data with a significantly wider PSF (i.e. more emitter overlap) than the training data.
Deep-STORM net2 and Deep-STORM net1 both yield very poor reconstructions, probably due to incompatible training, inevitable due to not knowing the exact imaging parameters. 
Altogether, LSPARCOM BT seems to be the best performer for this extremely challenging dataset.
In terms of runtime, SPARCOM ran for 37.6 sec, LSPARCOM reconstruction took 10.04 sec, and Deep-STORM recovery took 114.23 sec for $198$  $128\times128$ input frames. 

Figure 6(c) presents the results for another experimental tubulin sequence, composed of a sequence of 27,529 low-density frames. The original, low-density frames were used to create a ground-truth image, using the ThunderSTORM ImageJ plugin \cite{ThunderSTORM}. To create high-density frames for testing the performance of the methods in high temporal resolution, we summed each group of 50 frames to obtain 550 high-density frames. 
The imaging parameters of this dataset are unknown, but seem to be similar to those Deep-STORM net2 and LSPARCOM TU were trained on. 
SPARCOM with the explicit PSF, both versions of LSPARCOM, and both versions of Deep-STORM all yield good results, with the first two being more sparse at the expense of some false negative localizations, and the former being more noisy, similar to the ThunderSTORM result which we use as ground truth for reference. LSPARCOM TU also suffers from some false positive detections. 
The least sparse reconstruction was obtained by SPARCOM with a dirac delta PSF, with tubulins almost as thick as the diffraction limited image.
In terms of runtime, the SPARCOM reconstruction took 40.02 sec, the LSPARCOM reconstruction took 11.24 sec, and the Deep-STORM reconstruction took 307.87 sec for $550$  $128\times128$ input frames. 

To conclude the experimental portion of the results, LSPARCOM yields excellent results at a fraction of the runtime of the other methods, and provides a good reconstruction even when other methods fail. 
Note that LSPARCOM BT tends to reconstruct intersection areas as small arcs, a result of the inherently different training data, and may cause a ``chaining" affect when the PSF of the test data is significantly wider than that of the training data. 
Still, the decline in its performance due to this arcing effect is small relative to the decline in performance by Deep-STORM when dedicated training is not possible (e.g. since the imaging parameters are not known, such as in Fig. 6(b)), and to the decline in performance by SPARCOM when the PSF is unknown. 

\section{Conclusions and discussion}

The concept at the core of deep algorithm unfolding is using the algorithmic framework dictated by the underlying model to gain interpretability and domain knowledge, while inferring optimal parameters from the data. This places the unfolded network between the original algorithm (SPARCOM), which has low dependency on the type of input data as long as it fits the sparse prior (but is highly dependent on tuning the optimization parameter 
$\lambda$), and standard deep-learning based methods, such as ANNA-PALM \cite {ANNA-PALM} and Deep-STORM \cite {DEEP-STORM}, which have very strong dependencies on the type of input data. ANNA-PALM requires the test set to be identical to the training set in terms of structure \cite {ANNA-PALM}, and Deep-STORM - in terms of imaging parameters (mostly SNR and emitter density) \cite {DEEP-STORM}. As a data-driven method, LSPARCOM is also inevitably suitable for certain types of structures and imaging parameters more than others; this is the distribution we carved the solution for. Nevertheless, due to the significant algorithmic contribution to the network structure, the decline in reconstruction quality caused by diverging from this distribution is small relative to its standard-deep-learning counterparts. 
Fig. 6(b) is a good example of this robustness: while both versions of Deep-STORM yield reconstructions of very low quality such that they are irrelevant (due to mismatch between the training and test set), the exact same versions of LSPARCOM that were used to reconstruct all datasets, each trained from single FoV,  gave reasonable results, with no prior knowledge. The strong backbone of LSPARCOM is its reliance on an iterative method; this is evident from the results, which are visually similar to those obtained via SPARCOM with explicit knowledge of the PSF, but without the need for parameter selection or PSF prior knowledge. This algorithmic backbone also leads to a compact network, allowing faster inference than conventional deep learning techniques.

As explained in Section 4.4, the regularization parameter $\lambda$ used for training LSPARCOM sets the balance between the demand for accurate reconstruction in pixels that contain emitters, and the demand for minimal false detections. In spite of slight heuristics in this initial choice, it only has to be made once, for the training set. As was demonstrated in the Results section, the network generalizes well to different data types, using the exact same weights that were optimized for a loss using a specific value of $\lambda$ suitable for the training set, even when the $\lambda$ parameter used in SPARCOM differs significantly; in Figs. 2,4-6 the values of $\lambda$ (knowing the PSF) took five different values: 0.05, 0.1, 0.115, 0.13, and 0.25, whereas a single version of LSPARCOM led to comparable or superior results with no parameter adjustment.

As was demonstrated in the Results Section, 10 folds of LSPARCOM yield results which are comparable to running SPARCOM for 100 iterations with a carefully-chosen regularization parameter $\lambda$. It also requires no explicit knowledge of the PSF, but performs significantly better than the version of SPARCOM that assumes a dirac delta PSF. LSPARCOM takes after the general framework of SPARCOM, but replaces the need for explicit prior knowledge of the PSF, used in SPARCOM to calculate the measurement matrix, with learned filters. The mathematical justification for this replacement arises from replacing the original input to SPARCOM, which uses the information regarding the PSF to choose and weight elements of the covariance matrix, with a hard-choice of only the diagonal elements, yielding the variance matrix of the input movie, similar to the approach used in second-order SOFI with auto-cumulants and SUSHI  \cite{SOFI, SUSHI}. While in second-order auto-cumulant SOFI this leads to the final estimation of the super-resolved image, requiring no knowledge or estimation of the PSF, in SUSHI this image is the starting point to a sparse coding procedure. We used the resized variance matrix as the input to the unfolded algorithm, allowing approximate execution of the optimization procedure by learned convolution. LSPARCOM does theoretically compromise with respect to SPARCOM in the sense that it takes only the diagonal of the covariance matrix as input, thus not allowing any of the cross-correlation elements to contribute; yet the results prove that this compromise has negligible effect on the final reconstruction. 
In addition, LSPARCOM uses an adjustable threshold which is dependent on the local values, allowing a uniform reconstruction even in cases of location-dependent noise, thus offering another advantage over SPARCOM. 

The low performance of the version of SPARCOM that assumes a dirac delta PSF relative to LSPARCOM is not surprising, even though both practically use the variance of the temporal stack as input. This is because the assumption of a dirac delta PSF in SPARCOM affects not only the calculation of the input, but also the calculation of additional matrices used in the optimization procedure; this is substantially different than the case in LSPARCOM, where the PSF is considered and compensated for inside the network.

All the methods used for comparison allow high temporal resolution by using a relatively small number (few hundreds) of high emitter-density input frames. As demonstrated in Fig. 4(b), excellent reconstruction can even be obtained by LSPARCOM using as few as 25 frames. When implemented serially, LSPARCOM has the best execution times, with approximately 5 $\times$ improvement over SPARCOM and an order of magnitude improvement over Deep-STORM. While the former speedup is expected to remain somewhat constant, due to the innate improvement in running 10 folds rather than 100 iterations, the latter speedup is likely to be significantly reduced by using a designated environment inducing parallelism using the GPU. Still, the significant difference in model size - 9058 parameters in 10 folds of LSPARCOM vs. 1.3M trainable parameters in Deep-STORM - is expected to keep the runtime advantage with LSPARCOM. The structured framework in LSPARCOM along with the reduced model size (1166 trainable parameters considering the radial constraint) relative to Deep-STORM also yield more effective and efficient training, which prevents overfitting and generalizes well to various imaging parameters, as well as various geometries. This ability is crucial, as precise knowledge of the imaging parameters is not always available for creation of a designated training dataset. 

Along with its many advantages, LSPARCOM has several limitations. 
First, as was evident from the results, while the version of LSPARCOM trained on microtubules (BT) is incredibly robust, enabling good reconstruction of tubulins as well, it tends to reconstruct intersection areas as small arcs; this can be overcome by using a more relevant training set, as we did with the second version trained on tubulins (LSPARCOM TU), or by using a more versatile training set, consisting of more than one FoV.
Here, we intentionally used different data types for training (from a single FoV) and testing, to demonstrate the extreme generalization ability of the network.
Second, both versions of LSPARCOM are sometimes fragmented; this can be overcome by using a smoother regularizer (e.g. total variation) for deriving the iterative scheme used for unfolding, as shown in \cite{SPARCOM2}.
Third, samples of discrete nature, such as receptors spread upon a surface, tend to be less suitable to the classic SPARCOM reconstruction \cite {SPARCOM}, where the sparsity prior is on the spatial distribution of the fluorophores; this type of sample is more suitable for reconstruction considering sparsity in the wavelet domain \cite{SPARCOM2}. Since LSPARCOM was formulated based on the classic SPARCOM \cite{SPARCOM}, it too does not work well for such samples. Similarly to the fragmentation problem, this can be addressed by unfolding the appropriate extended version of SPARCOM \cite{SPARCOM2}, which we intend to do in future work. 
Fourth, due to the relative threshold parameters used in the proximal operator, LSPARCOM may yield false positive detections in empty patches. 
Finally, even though LSPARCOM has significant generalization ability for most imaging parameters, it is sensitive to the pixel size it was trained on - 100nm, similarly to other learning-based methods such as Deep-STORM. Its performance may also degrade for a significantly wider PSF than that of the training data, as in Fig. 6(b), causing a ``chaining" affect. However, altogether it has proven to be incredibly robust to the change of all other imaging parameters, including signal to noise ratio, and we believe that training on a more diverse training set will reduce many of these artifacts. 

To conclude, we presented a learned, convolutional version of SPARCOM, which relies on its framework for domain knowledge and surpasses it by incorporating data-driven parameters. We believe LSPARCOM will find broad use in single molecule localization microscopy of biological structures, potentially replacing its iterative counterpart and leading to robust, interpretable methods for live-cell imaging.
The full code, as well as a graphic user interface (GUI), is available online \cite{SPARCOM-CODE}.

\section*{Disclosures}
\noindent The authors declare no conflicts of interest.

\appendix

\paragraph{Appendix}

In the appendix, we provide additional figures enabling deeper understanding of the unfolding and training process: Fig. 7 expands upon the choice of input to the unfolding process (Section 4.1); Fig. 8 - upon the convolutional filters (Section 4.2); Fig. 9 - upon the activation function (Section 4.3); and Fig. 10 - upon the training of LSPARCOM TU (Section 5.1.1). In addition, Fig. 11 shows the inputs to the different algorithms, which were used to generate the figures in the sections 5.2 and 5.3, as well as the corresponding diffraction limited images. 

\begin{figure}[h] 
\centering
\includegraphics[scale=0.35]{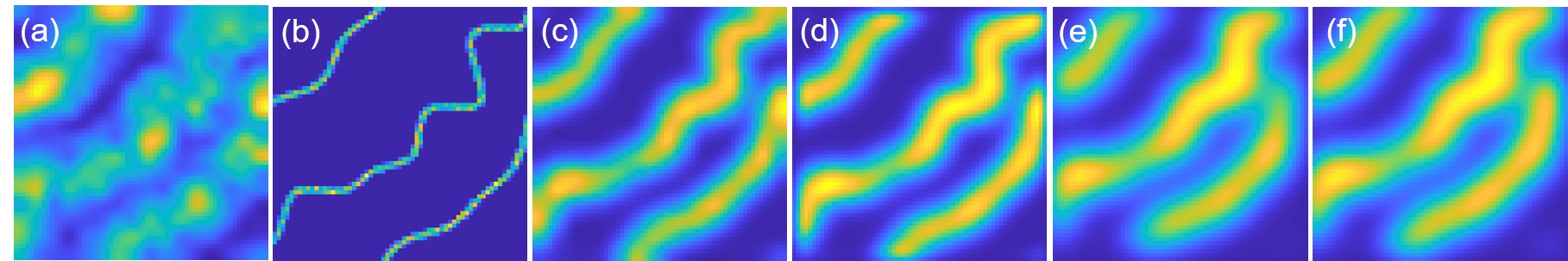}
\caption{\footnotesize Initial image estimation in various strategies, demonstrated for a synthetic dataset, simulating biological microtubules \cite{EPFL2015}.
(a) Diffraction limited image, obtained by summing the 361 frames of the simulation, interpolated to a $64\times64$ grid (from the original $16\times16$ grid).
(b) Ground truth image, obtained by summing the 361 localization frames.
(c) $\textbf{G}$, the temporal variance (matrix-shaped $\text{diag}\{\textbf{R}_y\}$) of the diffraction limited image stack, interpolated to a $64\times64$ grid (from the original $16\times16$ grid), equivalent to resizing a second-order SOFI image with auto-cumulants.
(d) $\textbf{v}$, calculated according to (9) considering a dirac delta PSF, reshaped as a matrix.
(e) $\textbf{v}$, calculated according to (9) considering the actual Gaussian PSF, reshaped as a matrix.
(f) $\textbf{v}$, calculated according to (15) considering the actual Gaussian PSF, reshaped as a matrix.}
\end{figure} 

\begin{figure}[h]  
\centering
\includegraphics[scale=0.4]{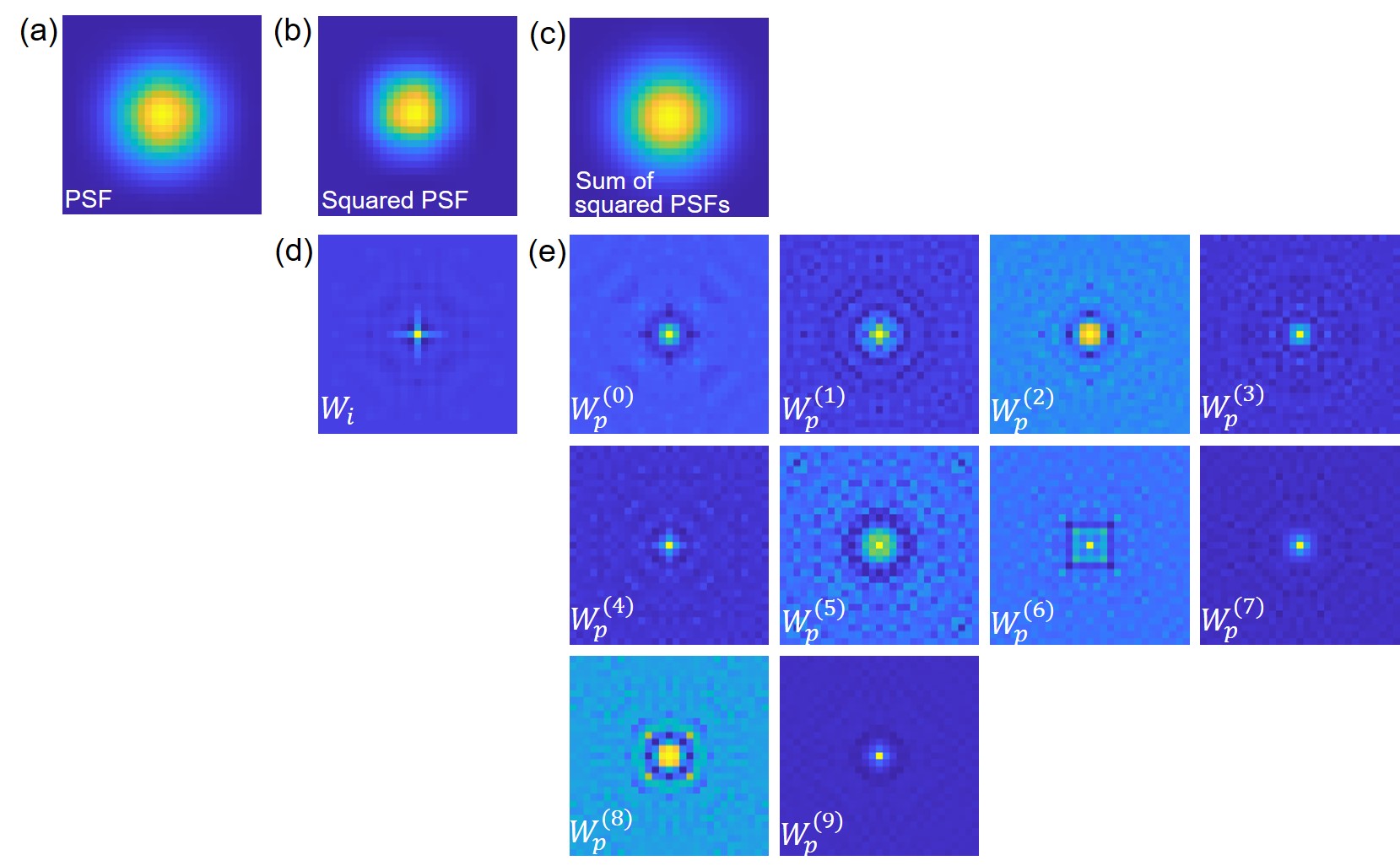}
\caption{\footnotesize Comparison of convolutional filters theoretical and learned values.
(a) The PSF of the training data, interpolated to the high resolution-grid and padded to a $29 \times 29$ window. 
(b) The squared PSF of the training data, interpolated to the high resolution-grid and padded to a $29 \times 29$ window (theoretical value of filter $W_i$). 
(c) The squared PSF of the training data, interpolated to the high resolution-grid and padded to a $29 \times 29$ window, shifted and weighted according to the low-resolution squared PSF (theoretical value of filters $W^{(k)}_p$, $k= 0,..9$).
(d) Learned value of $W_i$
(e) Learned values of $W^{(k)}_p$, $k= 0,..9$. }
\end{figure} 

\begin{figure}[h]  
\centering
\includegraphics[scale=0.5]{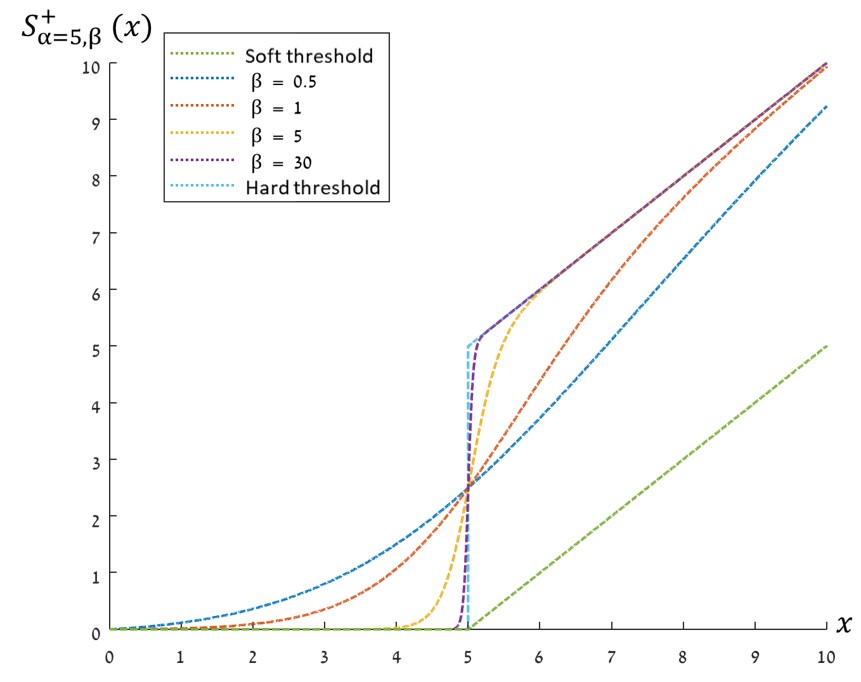}
\caption{\footnotesize The activation function used in LSPARCOM for $\alpha=5$ and various values of $\beta$, relative to a shifted ReLU (positive soft threshold) and a positive hard threshold. }
\end{figure} 

\begin{figure}[h]  
\centering
\includegraphics[scale=0.4]{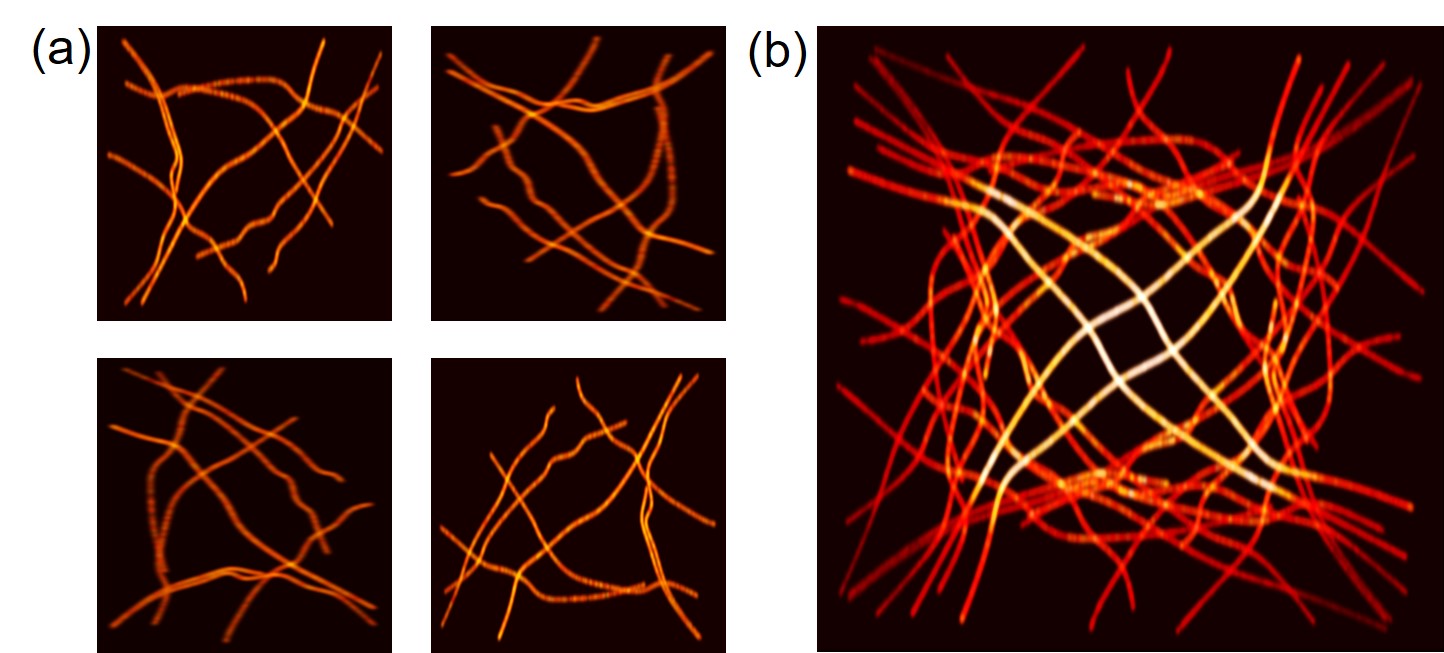}
\caption{\footnotesize FoV used for training LSPARCOM TU. The ground truth positions are based on the ones shown in Fig. 4, yet were generated by summing 4 such identical FoVs, rotated by either 0, 90, 180 or 270 degrees, to create a more dense FoV.}
\end{figure} 

\begin{figure}[h]  
\centering
\includegraphics[scale=0.4]{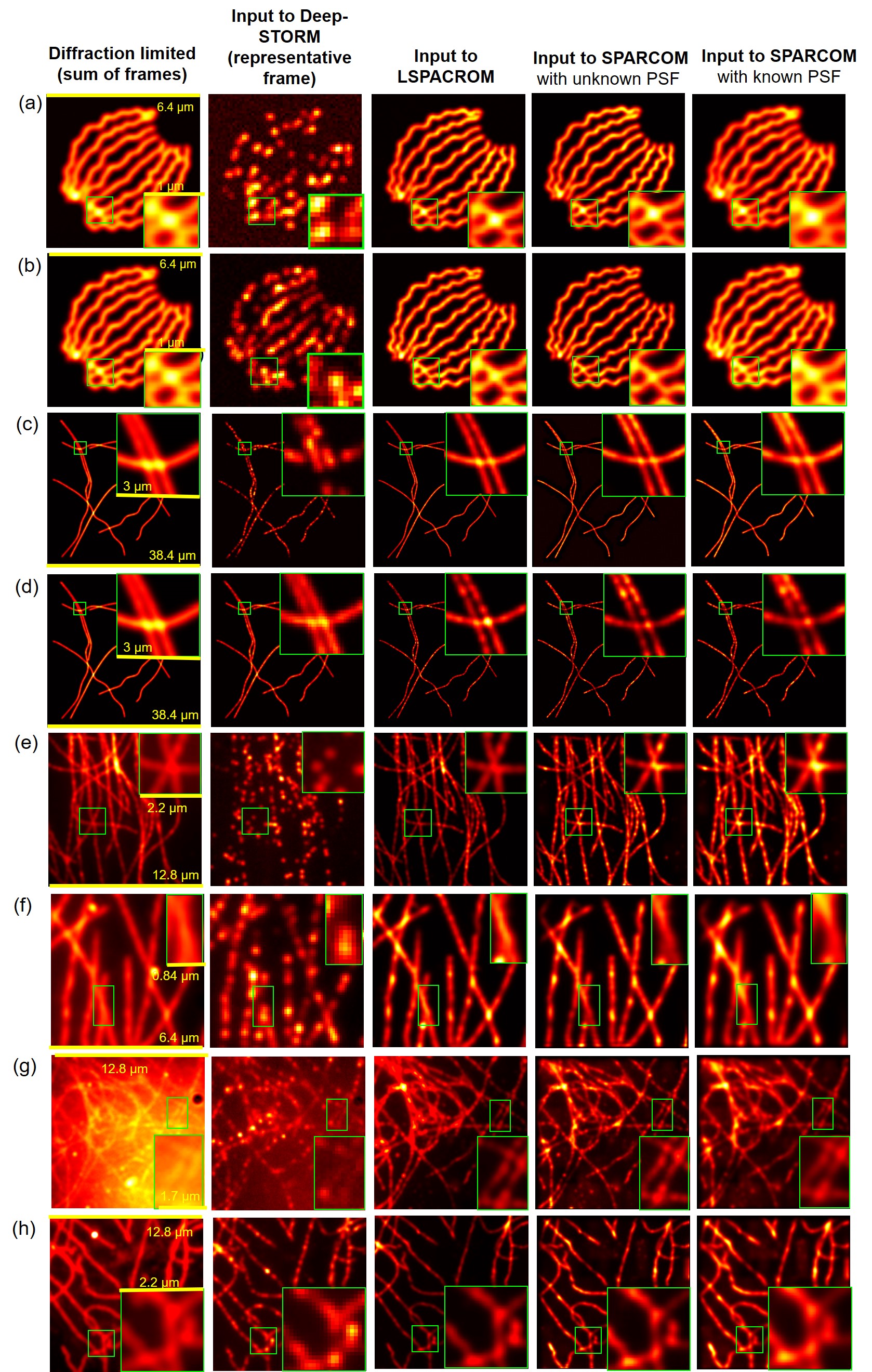}
\caption{\footnotesize Diffraction limited images and inputs to the different algorithms, corresponding to the various figures in the Results Section:
(a) Fig. 2(a).
(b) Fig. 2(b).
(c) Fig. 4(a).
(d) Fig. 4(b).
(e) Fig. 5.
(f) Fig. 6(a).
(g) Fig. 6(b).
(h) Fig. 6(c).
}
\end{figure} 

Figure 7 presents $\textbf{v}$ in the scenarios discussed in Section 4.1. 
Figure 7(a) displays the diffraction limited image and Fig. 7(b) displays the corresponding ground truth image. 
Figure 7(c) shows $\textbf{G}$, the temporal variance (matrix-shaped $\text{diag}\{\textbf{R}_y\}$) of the diffraction limited image stack, resized to the high resolution grid, equivalent to resizing a second-order SOFI image with auto-cumulants. 
Figure 7(d) illustrates $\textbf{v}$, calculated according to (9) considering a dirac delta PSF, reshaped as a matrix. 
As can be seen - Figures 7(c) and 7(d) are very similar (excluding boundary effect).
Finally, Figs. 7(e) and (f) display $\textbf{v}$, calculated according to (9) and (15), respectively, considering the actual Gaussian PSF, reshaped as matrices; as can be seen, they are very similar.
Note that even though Figs. 7 (e) and (f), which consider the actual PSF, seem further from the ground truth than Fig. 7 (d), which assumes the PSF to be a dirac delta function, the blurring is a crucial part of the reconstruction algorithm, which takes into account the underlying physical model, and thus needs to be incorporated into the unfolding procedure. 

Figure 8 compares between the theoretical values of the convolutional filters, as explained in Section 4.2, and their actual learned values. 
Figure 8(a) shows the PSF of the training data, interpolated to the high resolution-grid and padded to a $29 \times 29$ window. 
Figure 8(b) shows the squared PSF of the training data, interpolated to the high resolution-grid and padded to a $29 \times 29$ window. As explained in Section 4.2, this is the theoretical value of filter $W_i$, which convolving with is equivalent to multiplying by $\tilde{\textbf{A}}^T$ as in (15).
Figure 8(c) shows the squared PSF of the training data, interpolated to the high resolution-grid and padded to a $29 \times 29$ window, shifted and weighted according to the low-resolution squared PSF. As explained in Section 4.2, this is the theoretical value of filters $W^{(k)}_p$, $k= 0,..9$, which convolving with is equivalent to multiplying \textbf{M} by the high resolution vector-stacked image.
Figure 8(d) shows the learned value of $W_i$, and \mbox{Fig. 8(e)} shows the learned values of $W^{(k)}_p$, $k= 0,..9$. 
As can be seen, the learned values are inspired by the theoretical ones, but are not identical; this is to be expected, as LSPARCOM outperforms SPARCOM in terms of convergence speed and generalization ability, and also contains variable thresholds (as opposed to a constant one). 

Figure 9 shows $S^+_{\alpha=5,\beta}(x)$, as discussed in Section 4.3, for various values of $\beta$, relative to a shifted ReLU (positive soft threshold) and a positive hard threshold.

Figure 10 shows the FoV used for training the LSPARCOM TU net. The ground truth positions are based on the ones shown in Fig. 4; the final dense FoV was generated by summing 4 such identical FoVs, rotated by either 0, 90, 180 or 270 degrees.

Figure 11 shows the diffraction limited images as well as the inputs to the different algorithms, corresponding to the various figures in the Results section.
The first column to the left shows the diffraction limited images, which were obtained by resizing the sum of the input frames to the high resolution grid.
The second column shows a representative frame out of the input frames, since Deep-STORM operates on each frame individually, and then obtains the result by summing all reconstructions.  
The third column shows the input to LSPARCOM, which is actually the temporal variance of the input frames.
Finally, the fourth and fifth columns show the inputs to SPARCOM  (vector \textbf{v} from (9), reshaped as a matrix) without and with prior knowledge of the PSF, respectively. Note that due to memory restrictions, the inputs to SPARCOM were calculated by breaking the frames into overlapping patches, similarly to what is done for running the algorithm, as explained in Section 5.1.2. 

 \clearpage
 

\end{document}